%% file: RAA-2020-0451.tex
\documentclass{raa}
\usepackage{graphicx,times}
\usepackage{color}
\usepackage{natbib}
\usepackage{amssymb,amsmath}
 \usepackage[T1]{fontenc}
  \usepackage{aecompl}
\bibpunct{(}{)}{;}{a}{}{,}
\input{notation}

\begin{document}

   \title{A novel stellar spectrum denoising method based on deep Bayesian modeling}

 \volnopage{ {\bf 2016} Vol.\ {\bf X} No. {\bf XX}, 000--000}
   \setcounter{page}{1}

   \author{Xin Kang\inst{1}, Shiyuan He\inst{1}, and Yanxia Zhang\inst{2}}
   \institute{Institute of Statistics and Big Data, Renmin University of China, Beijing, 100872, P.R.China {\it heshiyuan@ruc.edu.cn}; \\
   \and CAS Key Laboratory of Optical Astronomy, National Astronomical
Observatories, Beijing, 100101, P.R.China {\it zyx@bao.ac.cn}\\
\vs \no
   {\small Received 2015 June 12; accepted }
}

\abstract{Spectrum denoising is an important procedure for large-scale spectroscopical surveys. This work proposes a novel stellar spectrum denoising method based on deep Bayesian modeling. The construction of our model includes a prior distribution for each stellar subclass, a spectrum generator and a flow-based noise model. Our method takes into account the noise correlation structure, and it is not susceptible to strong sky emission lines and cosmic rays. Moreover, it is able to naturally handle spectra with missing flux values without ad-hoc imputation. The proposed method is evaluated on real stellar spectra from the Sloan Digital Sky Survey (SDSS) with a comprehensive list of common stellar subclasses and compared to the standard denoising auto-encoder. Our denoising method demonstrates superior performance to the standard denoising auto-encoder, in respect of denoising quality and missing flux imputation. It may be potentially helpful in improving the accuracy of the classification and physical parameter measurement of stars when applying our method during data preprocessing.
}

   \authorrunning{X. Kang et al. }            
   \titlerunning{A novel stellar spectrum denoising method based on deep Bayesian modeling}  
   \maketitle

\section{Introduction}
With the rapid improvement of astronomical observation technology, modern large-scale sky surveys, such as the Sloan Digital Sky Survey (SDSS; \citealt{ahumada202016th}) and the Large Sky Area Multi-Object Fiber Spectroscopic Telescope (LAMOST or Guo Shoujing Telescope; \citealt{cui2012large}), provide unprecedented amount of astronomical data and enable us to explore our universe. The immense volume of astronomical data not only offers new opportunities but also brings challenges.

A fundamental data processing task is spectrum denoising when handling spectra. To clean astronomical spectra, wavelet is a standard tool. The wavelet shrinkage method applies the  wavelet transform to noisy observations, shrinks wavelet coefficients by some soft-thresholding or hard-thresholding rules, and takes the inverse wavelet transform to estimate the signal \citep{Donoho93nonlinearwavelet,Donoho94idealspatial,Donoho95adaptingto}. \cite{machado2013darth} developed a wavelet-based method for galaxy spectra and estimated their redshifts. Auto-encoder \citep{6796673,vincent2010stacked} is another popular denoising method in machine learning. Its success relies on  allowing only limited information to pass through a bottleneck for spectrum reconstruction. Through minimizing the loss objective function, the encoder learns a feed-forward latent representation of its input, and the decoder reconstructs the signal from the latent space.

Despite the success of these algorithms, there are still several unresolved issues. Standard wavelet method and auto-encoder require a complete spectrum as input. However, some spectral observations have missing flux values due to bad equipment conditions. Existing methods adopt an ad-hoc approach and directly impute the missing observations by some values (e.g. zero). In addition, the spectrum sometimes has a distorted shape and has wavelength-connection problem, because the full spectrum is combined from the blue and red channels. For some spectra, the two parts are misaligned. Lastly, the flux values sometimes get contaminated by strong night-sky emission lines or cosmic rays in each individual exposure. They induce bias to the existing denoising algorithms.

This work proposes a novel model to address the above problems. Based on deep Bayesian modeling, this paper puts forward a stellar spectrum denoising method, which not only denoises spectra but also recovers the defective spectra. Section~2 presents the description of used data. Section~3 introduces our proposed model. The application and experimental results of our model are indicated in Section~4. We summarize and conclude our paper in Section~5.

\section{Data}
\label{sec:data}

The Sloan Digital Sky Survey (SDSS; \citealt{ahumada202016th}) has been the most successful sky survey project in the history, which now provides images, optical spectra, infrared spectra, IFU spectra, stellar library spectra, and catalog data. Current data release is Data Release 16 (DR16). This work is conducted based on the stellar spectral observations from SDSS DR16. In this study, we select a comprehensive list of common stellar subclasses for model training and evaluation: O-type, B-type, A-type, F-type, G-type, K-type, M-type, Cataclysmic Variables (CVs), Carbon class, WD class (including CalciumWD, CarbonWD, WD, WDcooler, WDhotter).

Our proposed model will have several components: one prior distribution for each stellar subclass, a spectrum generator and a NoiseFlow observation model. See Section~\ref{sec:main} for more details.
For the training dataset of the spectrum generator, the top 200 spectra is selected among each stellar subclass sorted by the $r$-band signal-to-noise ratio (SNR).  Each selected spectrum is normalized to have unit absolute flux summation, and is interpolated to a fixed uniform grid with length of 2048 over the wavelength ranging from 4000$\mathring{A}$ to 9000$\mathring{A}$.  Meanwhile, the training dataset of the NoiseFlow observation model is prepared as follows. We extract the multiple-exposure data from the spectra with their SNR ranging from 20 to 30. For every exposure, we connect the red and blue parts of the spectrum, and apply the same interpolation and normalization as processing the training data. Then we subtract this composite spectrum from the average of multiple composite spectra. One subtracted spectrum from an exposure data becomes a training spectrum.

Test datasets also get prepared for model performance evaluation. We select an additional set of spectra with $r$-band SNR greater than 60. One test dataset consists of up to 300 spectra for each stellar subclass. We also select additional group of spectra whose $r$-band SNR is between 10 to 20 to extract new noise from their multiple-exposure data. The final test dataset is constructed by randomly adding these realistic noises to the spectra with high SNR. The goal for the denoising model is to reconstruct the original clean spectra with high SNR.

\section{The Denoising Model}
\label{sec:main}

\subsection{Deep Bayesian Modeling}
\label{sec:bayesian}
We now develop our proposed model for stellar spectrum denoising based on the basic Bayesian model~\eqref{eqn:basicBayesian:level1}--\eqref{eqn:basicBayesian:level2}. Suppose the signal spectrum of a star is $\vvs\in\bbR^{D}$. It is a $D$-dimensional unobserved vector, and we want to recover it from noisy observations. Modern astronomical surveys take multiple exposures to get several noisy observations $\vy_1,\cdots, \vy_n\in\bbR^{D}$ of the clean signal spectrum $\vvs$. We express the observations in a signal-plus-noise model
$$
\vy_i = \vvs + \vepsilon_{i},\ \text{ for }\  i = 1,\cdots, n
$$
where $\vepsilon_{i}$ is a $D$-dimensional noise vector. The noise could have complex correlation structure across pixels. Most of the time, only a single average spectrum is used, we can directly set $n = 1$ in the above. However, our method is more powerful and can exploit multiple-exposure data where only the red or blue part of the spectrum is recorded in each exposure. Our denoising framework is inspired by the following fundamental but powerful Bayesian model
\begin{align}
	\vy_1,\cdots,\vy_n |\vvs \sim & p(\vy | \vvs ),  \label{eqn:basicBayesian:level1}\\
	\vvs \sim & p(\vvs). \label{eqn:basicBayesian:level2}
\end{align}
In the above, $p(\vvs)$ is a prior density encoding the likelihood of the signal $\vvs$; $p(\vy | \vvs )$ is the probability density of the observed $\vy_i$ given the signal vector $\vvs$. Based on some state-of-the-art deep density estimation methods (see Section~\ref{sec:density}), we will construct an expressive prior $p(\vvs)$ and an observation model $p(\vy | \vvs )$ by neural networks. Given the trained model and the observations $\vy_1,\cdots, \vy_n$, the true signal vector $\vvs$ can be inferred from the posterior distribution $p(\vvs| \vy_1,\cdots,\vy_n)$.

Several additional adjustments of the model are necessary. It is not straightforward to model a clean spectrum $\vvs\in\bbR^{D}$ in a high dimensional space $\bbR^D$. To effectively construct a model for the signal spectrum, we exploit that, for a collection of astronomical spectra, the signal vectors $\vvs$ of various stars typically reside over a low dimensional manifold. This intrinsically low-dimensional structure can be effectively employed for spectrum data analysis. For example, \cite{lawlor2016mapping} used a local non-linear dimension reduction technique to discover the manifold structure. The low dimensional nature of the data implies that all clean spectra can be represented by a variable in a low dimensional space. Denote this low-dimensional variable by $\vz\in\bbR^{L}$, and  we can construct a mapping $\sG$ such that the signal $\vvs = \sG(\vz)$ is the mapped value of $\vz$. Based on this mapping, the  prior over the signal $\vvs $ can be directly expressed as a prior over the latent space $p(\vz)$. More specifically, we will take the stellar class $C$ into account and construct the prior $p(\vvs|C)$  conditional on each stellar subclass. The final hierarchical model is summarized as below.
\begin{align}
	\vy_1,\cdots,\vy_n |\,\vvs\, \sim & p(\vy | \vvs ),  \label{eqn:model:level1}\\
	\vvs = & \sG(\vz), \label{eqn:model:level12}\\
	\vz |C \sim & p(\vz | C), \label{eqn:model:level2} \\
	C \sim&  p(C).  \label{eqn:model:level3}
\end{align}
In this model, the class label variable $C$ has a uniform prior distribution over all stellar subclasses. The construction and training of the latent priors $p(\vz | C)$ and the generator function $\sG$ will be addressed in Section~\ref{sec:main:priormodel}. The observation model $p(\vy | \vvs )$ will be discussed in Section~\ref{sec:main:obsmodel}. The final model training and denoising workflow will be summarized in Section~\ref{sec:main:workflow}.
	
Given our trained model in~\eqref{eqn:model:level1}--\eqref{eqn:model:level3},
the posterior distribution of the latent variable $\vz$ and the class label $C$ is proportional to the joint distribution
\begin{align*}
	p(\vz, C | \vy_1,\cdots, \vy_n ) \propto \prod_{j=1}^{n} p(\vy_j | \sG(\vz)) \times p(\vz | C) \times p(C).
\end{align*}
Monte carlo Markov chain (MCMC) methods can be applied to draw samples from the posterior distribution. However, MCMC technique is known to be computationally expensive and not scalable to large datasets. For large scale modern astronomical surveys, we can adopt the faster maximum-a-posterior (MAP) estimation. In this way, the latent variables $\vz$ and the class label $C$ are found by
\begin{align}
	\hat{\vz}, \hat{C} = \argmax_{\vz, C} \Biggl\{\sum_{j=1}^{n} \log  p(\vy_j | \sG(\vz)) +
	\log p(\vz | C) + \log p(C) \Biggr\}. \label{eqn:bayesianMAP}
\end{align}
With the computed MAP estimation $\hat{\vz}$, the cleaned and denoised spectra are given by $\hat{\vvs} = \sG(\hat{\vz} )$.

\subsection{Background on Deep Density Estimation}
\label{sec:density}

This subsection reviews some works on deep density estimation. These works serve as the basis for building our deep Bayesian denoising model.
Suppose $\vx\in\bbR^D$ represent a $D$-dimensional observation, for which we want to estimate its distribution. In the deep learning literature, most density estimation methods are based on the intuition that we can transform a simple base density $\pi_\vz(\vz)$ into a more complex one $p(\vx)$ via an invertible differentiable transform $\vx = f(\vz)$. The function $f$ is parameterized by a neural network to achieve flexible and adaptive transformation. By the basic formula of density transformation, it holds that
\begin{equation} \label{eqn:densitytransform}
	p(\vx) = \pi_{\vz}(f^{-1}(\vx))\left|\det\left(\frac{\partial f^{-1}}{\partial \vx}\right)\right|.
\end{equation}

Several methods have been proposed for deep density estimation. The approach of  normalizing flows \citep{dinh2014nice} chooses $f$ as a sequence of composite functions, i.e.,
$f = f_1\circ f_2 \circ \cdots \circ f_K$. In this way, $p(\vx)$ can be regarded as an invertible and differentiable transformation $f$ of a base density $\pi_{\vz}(\vz)$. The base density can be simple multivariate Gaussian distribution. The relationship between the observation data $\vx$ and latent variable $\vz$ can be represented as $\vx \stackrel{f_1}\longleftrightarrow \vh_1\stackrel{f_2}\longleftrightarrow \vh_2 \cdots \stackrel{f_K}\longleftrightarrow \vz$ via a stack of hidden variables. Under the invertible (or  bijective) assumption of $f$, $\vz$ can be calculated as $\vz = f^{-1}(\vx)$ given $\vx$. Based on~\eqref{eqn:densitytransform}, the log probability density can be directly computed as
\begin{equation}
	\label{3}
	\begin{aligned}
		\log p(\vx) &= \log\pi_{\vz}(\vz) + \log\left\lvert\det\left(\frac{d\vz}{d\vx}\right)\right\lvert,\\
		&=\log\pi_{\vz}(\vz) + \sum_{i=1}^K\log\left\lvert\det\left(\frac{d\vh_i}{d\vh_{i-1}}\right)\right\lvert.
	\end{aligned}
\end{equation}
where the scalar value $\log\left\lvert\det(d\vh_i/d\vh_{i-1})\right\lvert$ is the logarithm of the absolute value of the determinant of the Jacobian matrix $(d\vh_i/d\vh_{i-1})$, also called log-determinant. The series of transformations are required to be easily invertible and the log-determinant should be easy to be computed. \cite{rezende2016variational} devised planar and radial flow as basic blocks for $f$. NICE \citep{dinh2014nice} adapts additive coupling layers to form a normalizing flow, and  its successor Real NVP \citep{dinh2017density} extends the transformation by stacking affine coupling layers. They all have a tractable triangular Jacobian matrix for the bijective mapping.

Autoregressive flow is another popular and tractable approach to density estimation. It factorizes the joint density as a product of conditional densities $p(\vx) = \prod_dp(x_d|\vx_{1:d-1})$ via the chain rule of probability \citep{uria2016neural}. This factorization makes the Jacobian tractable as the Jacobian becomes a triangular matrix. The inverse autoregressive flow \citep[IAF,][]{kingma2017improving} and the masked autoregressive flow \cite[MAF,][]{papamakarios2018masked} take this approach.
Under this approach, the prediction of current value depends on all of its past values, which is referred to as the autoregressive property. They use independent standard Gaussian distributions as the base density. The mean and variance of
$x_d$ are functions of the preceding observation vector $\vx_{1:d-1}$ or the preceding random numbers. They both use MADE \citep{germain2015made} as their basic building blocks for the function mapping.

\subsection{The Generator and Latent Prior}
\label{sec:main:priormodel}

\begin{figure}
	\centering
	\begin{minipage}{0.49\linewidth}
		\includegraphics[width=0.88\textwidth]{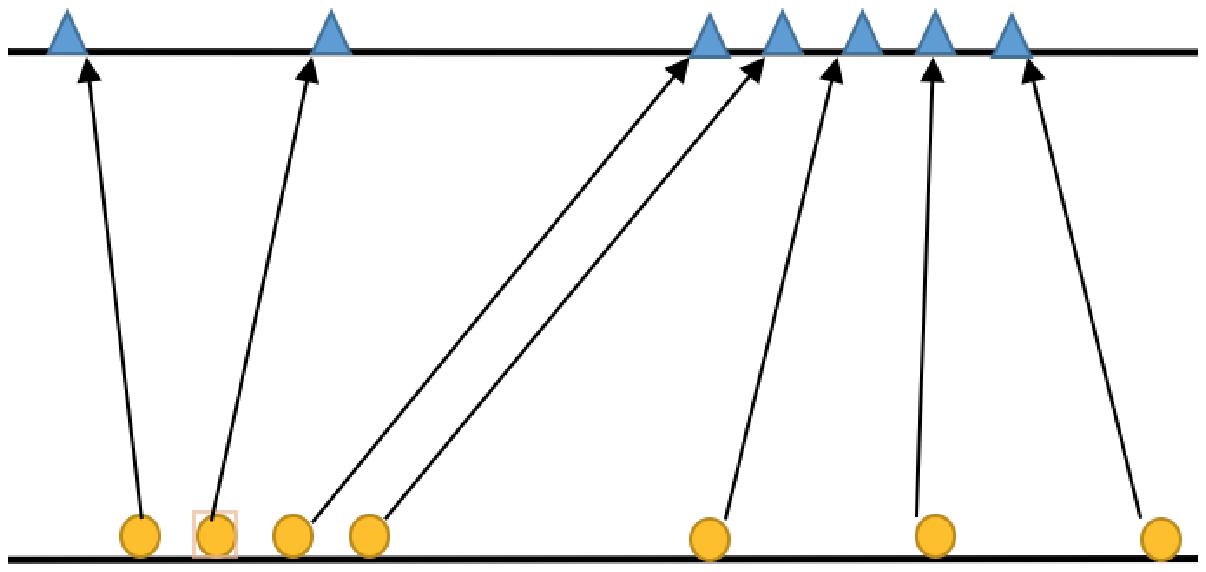}
	\end{minipage}
	\begin{minipage}{0.49\linewidth}
		\includegraphics[width=0.95\textwidth]{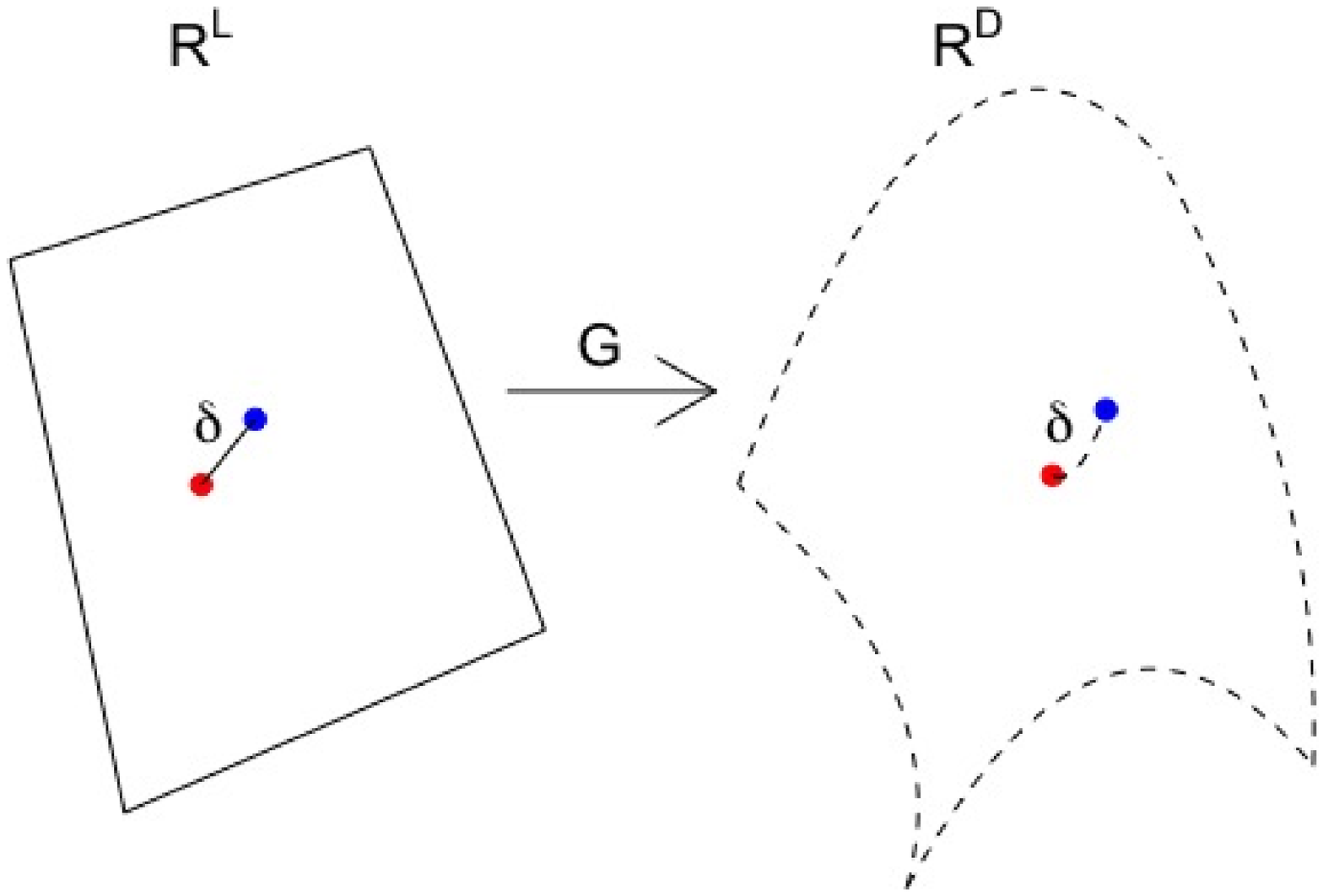}
	\end{minipage}
	\caption{The left subfigure illustrates a distorted mapping where the density estimation is affected. The right subfigure illustrates a locally isometric mapping where the local distance is preserved. Our generator $\sG$ implements this isometry property to avoid distortion in the latent space.}
	\label{fig:toy:distrot}
\end{figure}

We first detail our spectrum generator $\sG$ and the prior density $p(\vz|C)$ for the latent variable. The generator is obtained from the auto-encoder framework, but with an additional local isometry constraint. In the standard auto-encoder framework, an encoder $\sE$ maps an observation $\vy$ to the latent $\vz = \sE(\vy)$, and then maps $\vz\in \bbR^L$ back to the original high dimensional space $\bbR^D$ by the generator (decoder) $\sG$. The training target is to minimize the reconstruction loss
$$
\min_{\sG,\sE} \,\Expect_{\vy}  \| \vy - \sG(\sE(\vy)) \|^2 .
$$
The standard convolutional auto-encoder architecture can be specified for $\sE$ and $\sG$. However, the generator $\sG$ obtained hereby could create distortion in the latent space $\bbR^L$, affecting density estimation. The left subfigure of Figure~\ref{fig:toy:distrot} shows an example of distorted mapping, where some low-density points (on the bottom right) are mapped to a high-density region (on the top right). To avoid the issue,
we construct a \textit{locally isometric} mapping such that $\sG$ preserves the distance between samples in the latent space $\bbR^{L}$. In other words, it holds that
$$
\|\sG(\vz) - \sG(\vz') \| \approx \| \vz - \vz'\|.
$$
where $\vz,\vz' \in \bbR^{L}$  is a pair of latent variables satisfying $ \| \vz - \vz'\| \le \delta$ for some $\delta$. The local isometry property is illustrated in the right subfigure of Figure~\ref{fig:toy:distrot}, where the distance between the red and blue points is preserved under the mapping $\sG$. The isometry property allows us to obtain the spectrum density $p(\vvs)$ by directly accessing the low dimensional $\vz$.  It approximately holds that $p(\vvs) = p(\sG(\vz)) = p(\vz)$. This helps us to approach the density estimation and effectively avoid the computation of $\det \frac{\partial f^{-1}}{\partial \vx}$ for a complex transformation function $f$ in~\eqref{eqn:densitytransform}.

With the additional local isometry constraint, the objective function to train the generator becomes
\begin{equation}
\min_{\sG,\sE}\, \Expect_{\vy,\vdelta}
\Big\{ \big[\|\sG(\vz) - \sG(\vz+\vdelta) \| - \delta\big]^2
+ \| \vy - \sG(\sE(\vy)) \|^2 \Big\}. \label{eqn:generatorObj}
\end{equation}
In the above, the latent value is  $\vz = \sE(\vy)$. The perturbation $\vdelta$ is a random variable, drawn from a uniform distribution over the sphere of radius $\delta$ in $\bbR^L$. Similar loss functions had been employed in the works \citep{geng2020uniform,atzmon2020isometric} to train auto-encoder to learn the manifold structure.

After training $\sE,\sG$ based on~\eqref{eqn:generatorObj}, we can compute the latent variables $\vz$ for all training samples. Then, based on this collection of latent variables, a kernel density estimator (KDE) is deployed over $\vz$ for the training samples of each stellar subclass $C$. This helps us to obtain $p(\vz| C)$ for each subclass $C$.

\subsection{The NoiseFlow Observation Model}
\label{sec:main:obsmodel}

This subsection constructs the observation model $p(\vy|\vvs)$ for $\vy_n$ given the true signal $\vvs = \sG(\vz)$. Recall that we have used the additive noise model $\vepsilon_n = \vy_n - \vvs$. Our goal is equivalent to construct a noise density model $p(\vepsilon_{n})$ and set
$p(\vy_n |\vvs) = p(\vepsilon_{n})$. The density model is built upon the idea of
\cite{papamakarios2018masked} such that the observation model has an auto-regressive structure. Compared with their model, our model only depends on a local group of pixels within an observation, is more parsimonious in parameterization, and focuses on extracting local correlation structure of the noise.

Suppose the noise vector is written as $\vepsilon_n = (\epsilon_{n1},\epsilon_{n2},\cdots, \epsilon_{nK})$. For $d =K+1,\cdots, D$, the noise value $\epsilon_{nd}$ for the $d$-th pixel depends on its immediate $K$ preceding pixels via
 \begin{equation} \label{eqn:noiseflow}
	p(\epsilon_{nd}|\vepsilon_{n,(d-K):(d-1)}) = \mathcal{N}(\epsilon_{nd} |\mu_{nd}, (\exp\alpha_{nd})^2).
\end{equation}
where $\mu_{nd} = f_{\mu}(\vepsilon_{n,(d-K):(d-1)} )$, $\alpha_{nd} = f_{\alpha}(\vepsilon_{n,(d-K):(d-1)} )$, and $f_{\mu}$ and $f_{\alpha}$ are two functions expressed by neural networks. The architecture of  $f_{\mu}$ and $f_{\alpha}$ will be specified later. The two functions $f_{\mu}$ and $f_{\alpha}$ determine the mean and the standard deviation for the noise $\epsilon_{nd}$ at the $d$-th pixel. In particular, we have
\begin{equation}
	\epsilon_{nd} = \xi_{nd}\exp(\alpha_{nd}) + \mu_{nd}. 	\label{eqn:xi2epsilon}
\end{equation}
where  $\xi_{nd} \sim \mathcal{N}(0,1)$ follows the standard Gaussian distribution. Equivalently, the random variable can be expressed by the following inverse expression
\begin{equation}
	\xi_{nd} = (\epsilon_{nd}- \mu_{nd})\exp(-\alpha_{nd}). 	\label{eqn:epsilon2xi}
\end{equation}
The above specifies the local dependence structure for $d=K+1,\cdots, D$. For the first $K$ pixels, there are not enough preceding pixels for us to determine the conditional likelihood~\eqref{eqn:noiseflow}. Instead, for the first $K$ pixels, the noise $\epsilon_{nd}$  is imposed to follow a univariate Gaussian distribution with a fixed mean $\mu_{d}$ and a fixed standard deviation $\sigma_{d} = \exp(-\alpha_{d})$. In summary, our model parameters to be trained include: the scalar values $\mu_{d},\alpha_{d}$ for $d=1,\cdots, K$; and two neural network functions $f_{\mu}$ and $f_{\alpha}$.

As $\mu_{nd}$ and $\alpha_{nd}$ depend on $\vepsilon_{n,(d-K):(d-1)}$ by design,  the Jacobian in density transformation~\eqref{eqn:densitytransform} is an upper triangular matrix. As a result, the absolute value of the determinant can be easily computed as
\begin{equation}
	\log\Big|\det \frac{\partial f^{-1}}{\partial \vepsilon} \Big| = -\sum_{d=K+1}^D\alpha_{nd}.
\end{equation}
It follows that the log density of the observation model becomes
\begin{align}
\log p(\vy_n| \vvs) = & p(\vepsilon_{n}) \nonumber \\
=&\sum_{d=1}^{K} \log p(\epsilon_{nd}) +
\sum_{d=K+1}^{D} \log p(\epsilon_{nd}|\vepsilon_{n,(d-K):(d-1)})\nonumber  \\
 = &  \sum_{d=1}^{K}
\Big[- (1/2)\exp(-2\alpha_{d})(\epsilon_{nd} - \mu_d)^2  - \alpha_{d}\Big] \nonumber \\
&\quad +\sum_{d=K+1}^{D}
\Big[- (1/2)\exp(-2\alpha_d)(\epsilon_{nd} - \mu_{nd})^2  - \alpha_{nd}\Big] + \mathrm{const}.\nonumber
\end{align}
The last equality holds up to some irrelevant constant $\mathrm{const}$.

 \begin{figure}[t]
 	\centering
 	\includegraphics[scale=0.35]{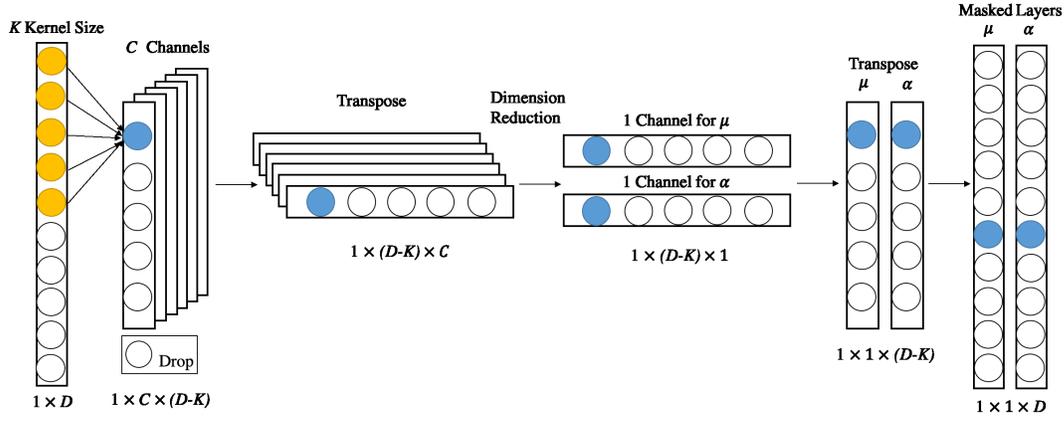}
 	\caption{The neural network architecture of our observation model. The blue pixel only depends on the immediately preceding five yellow pixels of the input spectrum. This proposed architecture is more parsimonious in parameterization and focuses on extracting local correlation structure of the noise.}
 	\label{fig:noiseflow}
 \end{figure}

For our observation model, we develop a neural network architecture for local noise feature extraction, as shown in {Figure~\ref{fig:noiseflow}}. In the first layer, local features are extracted by an one-dimensional convolution layer. This convolution layer has one input channel and $C$ output channels, kernel size $K$, one stride and zero padding. Its output $\vh$ is of dimension $N\times C\times (D-K+1)$, where $N$ is the mini-batch sample size. In accordance of the aggressive structure, we drop one redundant column (the first column) in the third dimension of the hidden state $\vh^1$, such that its dimension becomes $N\times C\times (D-K)$.  We then take a transpose to exchange the second dimension with the third dimension. The resulting $\vh$ is of dimension $N\times (D-K)\times C$. In this way, for the $d$-th pixel ($d=K+1,\cdots, D$), we get a $C$-dimensional feature vector for it.

After that, $\vh$ is used as the input of the following $L-1$ linear hidden layers with RELU, to sequentially reduce the dimension of hidden variables from $N\times (D-K)\times C$ to $N\times (D-K)\times C'$ for some $C'<C$. In the sequel, there are two separate linear hidden layers mapping from $N\times (D-K)\times C'$ to $N\times (D-K)\times 1$, one is for $\vmu$ and the other is for $\valpha$, and we transpose $\vmu$ and $\valpha$ back to $N\times1\times(D-K)$. At this moment, vector $\vmu$ and $\valpha$ contain the mean and standard deviation information for $\epsilon_{nd}$ with $d=K+1,\cdots, D$.
As for the first $K$ pixels, their scalar mean and standard deviation values ($\mu_{d}$, $\alpha_{d}$ for $d=1,\cdots, K$) get included via the final masked linear layers.

\begin{figure}[t]
	\centering
\includegraphics[scale=0.45]{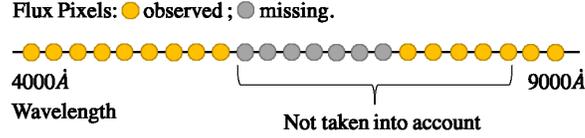}
\caption{The masked likelihood for a partially observed spectrum. The yellow pixels represent the observed pixel values and the grey pixels correspond to the missing ones. The likelihood of a pixel is accounted if and only if the $d$-th pixel and its immediate $K$ preceding pixels are observed. Our observation model can naturally deal with spectra with missing values. }
	\label{fig:masked}
\end{figure}

Moreover, the observation model developed hereby can naturally handle a spectrum with missing flux values due to bad pixels. We can create a mask vector $\vm_n$ such that
$m_{nd} = 1$ if all of $y_{n,d-K}, \cdots, y_{n,d-1},y_{n,d}$ are observed, and $m_{nd} = 0$ otherwise.
The observation log-likelihood for $\vy_n$ becomes
\begin{equation} \label{eqn:maskedlikelihood}
\log p(\vy_n| \vvs) = \sum_{d=1}^{K} m_{nd}\log p(\epsilon_{nd}) +
\sum_{d=K+1}^{D} m_{nd}\log p(\epsilon_{nd}|\vepsilon_{n,(d-K):(d-1)}).
\end{equation}
In other words, the likelihood of the $d$-th pixel is taken into account if and only if the $d$-th pixel and its immediate $K$ preceding pixels are observed, which is shown in {Figure~\ref{fig:masked}}. The masked likelihood allows us to deal with partially observed spectrum without resorting to ad-hoc missing value imputation.

\subsection{Modeling Workflow}
\label{sec:main:workflow}

\begin{figure}
	\centering
    \includegraphics[scale=0.4]{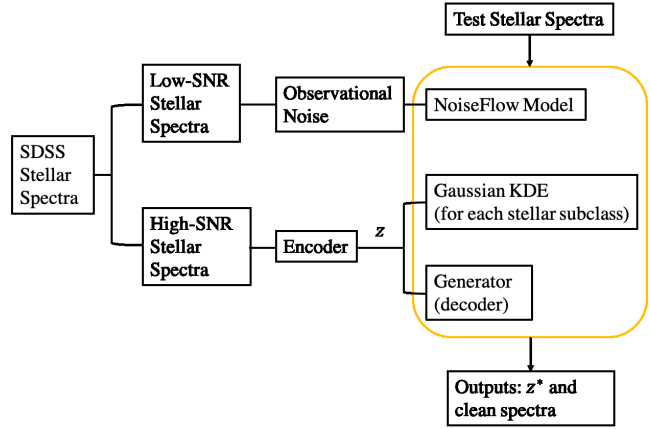}
    \caption{The modeling workflow of our method. The core Bayesian model in the yellow box consists of three parts: one spectrum generator, one Gaussian KDE for each stellar subclass and one NoiseFlow model. Spectra with various levels of SNR are supplied for the training of different model components. With iterative optimization, it outputs the corresponding latent variables and the denoised spectra.}
	\label{fig:architecture}
\end{figure}

The main workflow of our model is summarized in {Figure~\ref{fig:architecture}}.
In order to train and apply our proposed model, basically four main steps are involved:
\begin{enumerate}
	\item Train an encoder $\sE$ and a locally isometric generator $\sG$ based on a collection of high-SNR optical stellar spectra.
	\item Use Gaussian kernel density estimation (Gaussian KDE) to estimate the prior distribution of the latent variable $\vz$ for each stellar subclass obtained from the encoder trained in the first step.
	\item Train a NoiseFlow observation model with the observational noise extracted from low-SNR optical stellar spectra.
	\item For each test spectrum, use the Gaussian KDE, the NoiseFlow model and the generator $\sG$ trained in the first three steps to obtain its corresponding latent variable and its clean and complete spectrum with iterative optimization of~\eqref{eqn:bayesianMAP}.
\end{enumerate}

\section{Results}
This section compares our method with the convolutional denoising auto-encoder based on two test datasets. In particular, two tasks are created for the purpose of spectrum denoising and missing flux imputation. The experiment is conducted based on the stellar spectral observations introduced in Section~\ref{sec:data}.

\begin{figure}
	\centering
	\includegraphics[scale=0.44]{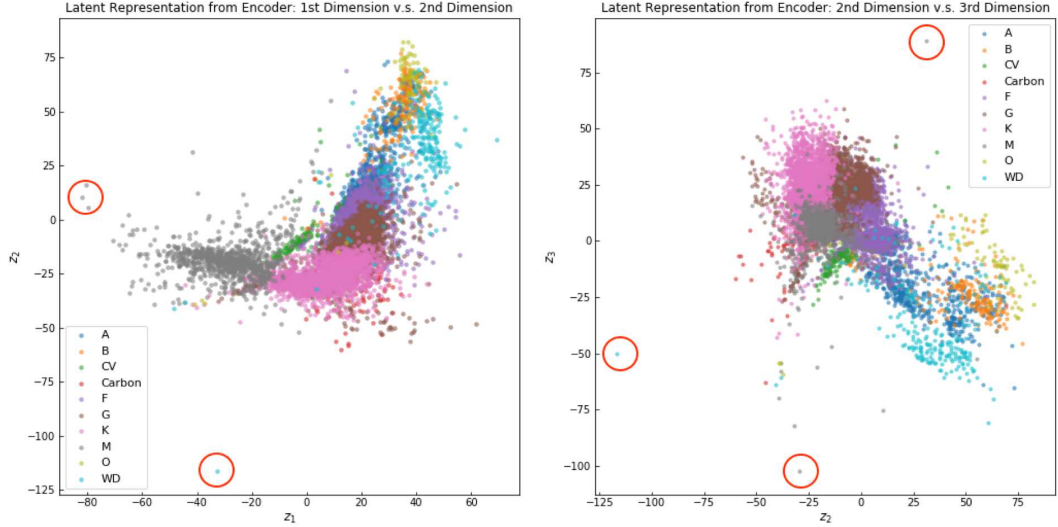}
	\caption{The three dimensional space for the latent variable $\vz$ outputted by the encoder. The horizontal axis in left subfigure is $z_1$ and the vertical axis is $z_2$. The horizontal axis in right subfigure is $z_2$ and the vertical axis is $z_3$. It is evident that observations from the same stellar subclass form a cluster in the latent space. The red circles indicate the outlier spectra. This learned latent space informs the latent variable structure for most stellar observations.}
	\label{fig:latent}
\end{figure}

The proposed model is trained over the training dataset of Section~\ref{sec:data}.  Recall that each training spectrum has been interpolated over a grid with size of $D=2048$, and we will set the latent space dimension as $L=3$. This training dataset is supplied to Equation~\eqref{eqn:generatorObj} to train the generator $\sG$ and the encoder $\sE$ with stochastic gradient algorithm. The encoder learns a latent representation $\vz \in \mathbb{R}^3$ for each training spectrum. Figure~\ref{fig:latent} shows the scatterplot of the latent variables for the training dataset. The points are colored according to the stellar subclasses. It is evident that observations from the same stellar subclass form a cluster. The latent space can also help us to detect outlier spectra, such as those inside the red circles indicated in {Figure~\ref{fig:latent}}. This learned latent space informs us the latent variable $\vz$ structure for most stellar observations. Therefore, we can use a Gaussian kernel density estimation for each of ten stellar subclasses to get $p(\vz|C)$.

For comparison, the convolutional auto-encoder gets trained based on a larger dataset. The dataset contains spectra both with high SNR and low SNR. Besides, for a fair comparison, the  convolution neural network shares the same architecture (e.g. the same hidden layers and the same transposed convolution layers) as our generator $\sG$.

\subsection{Spectrum Denoising}
\label{sec:test:noise1}

To test the model performance, we randomly select an independent group of spectra with high SNR for each stellar subclass, as described in Section~\ref{sec:data}. These selected spectra are regarded as the ground truth that the model endeavors to predict. Then we extract another noisy dataset from an independent group of stellar spectra with SNR ranging from 10 to 20. The noise is randomly sampled and added to the above clean test spectra. These constitute our benchmark test dataset for method comparison. Each stellar subclass is created with 300 test spectra.

\begin{figure}
	\centering
	\includegraphics[scale=0.4]{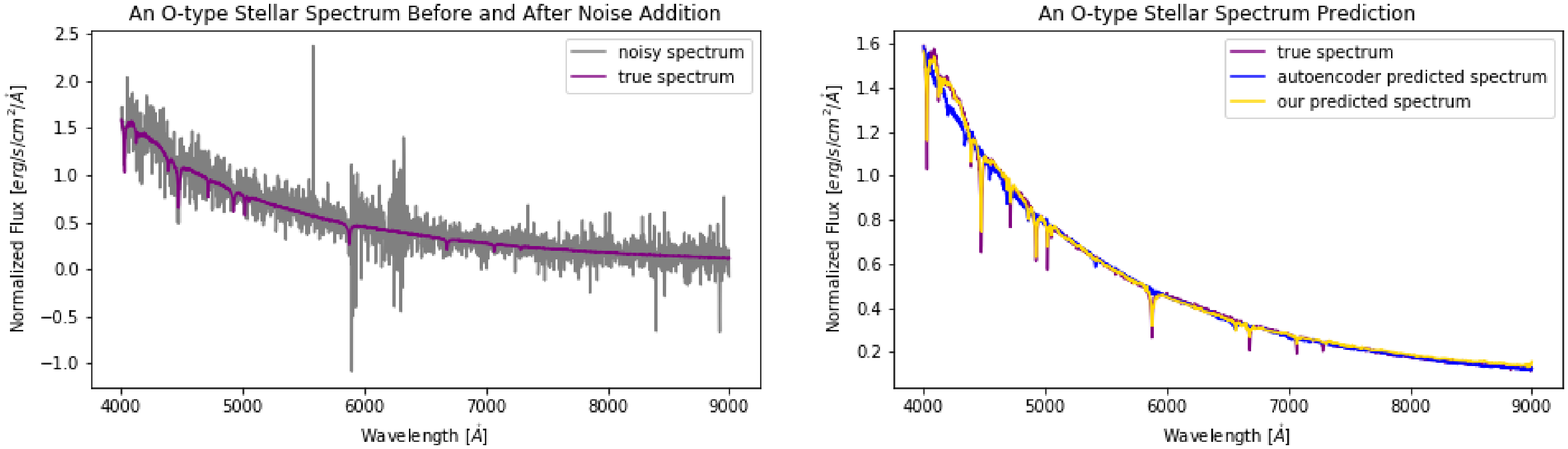}
	\includegraphics[scale=0.4]{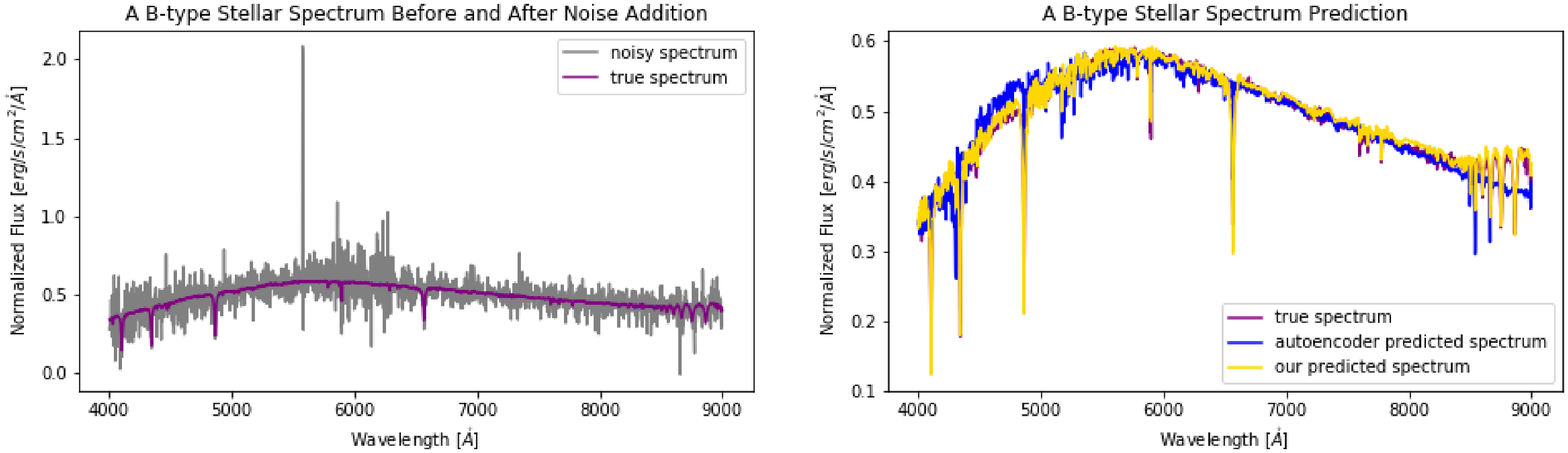}
	\includegraphics[scale=0.4]{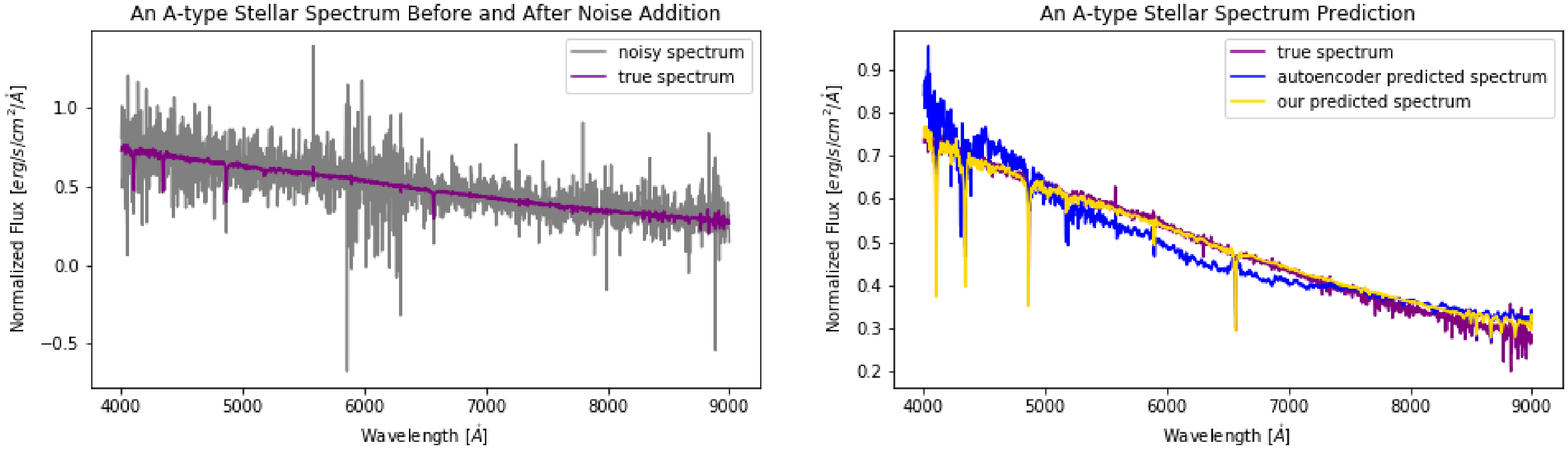}
	\includegraphics[scale=0.4]{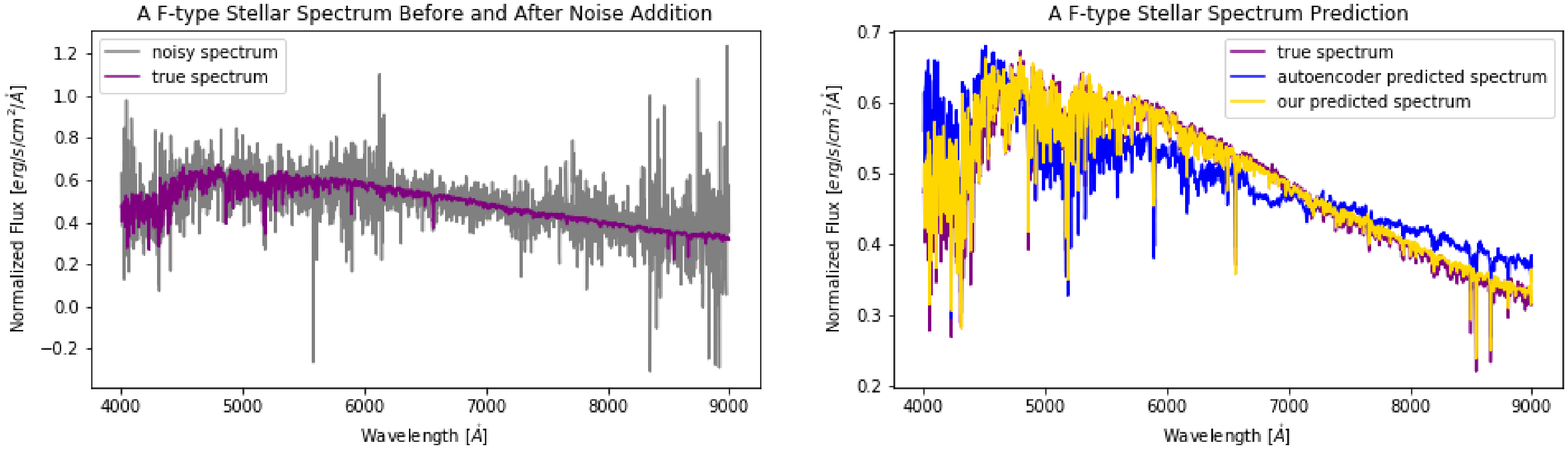}
	\includegraphics[scale=0.4]{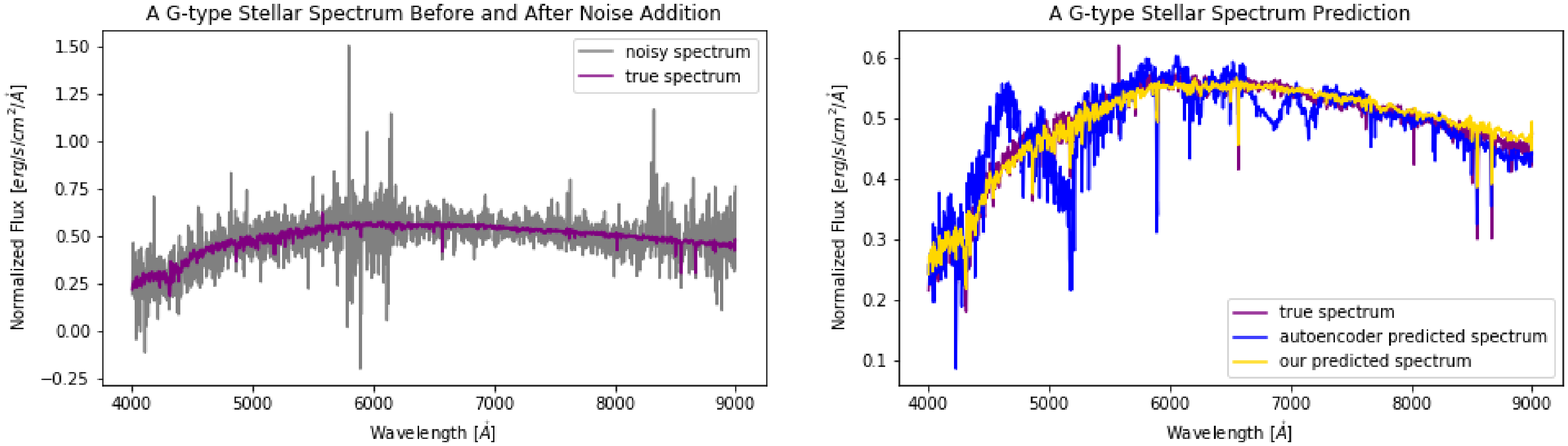}
	\caption{The five rows show examples of stellar spectra denoised by the convolutional denoising auto-encoder and our method for the stellar subclasses O, B, A, F, G, respectively. The left column shows the true spectrum and its synthetic counterpart with noise. The purple curve is the clean spectrum that both denoising algorithms try to recover, and the grey curve is the  spectrum added with  noise. The  spectra denoised by both algorithms are compared  in the right column. The denoised spectrum from our model (yellow curve) is much closer to the purple clean spectrum than the standard auto-encoder result (blue curve).
	\label{fig:testing1}}
\end{figure}

\begin{figure}
	\centering
	\includegraphics[scale=0.4]{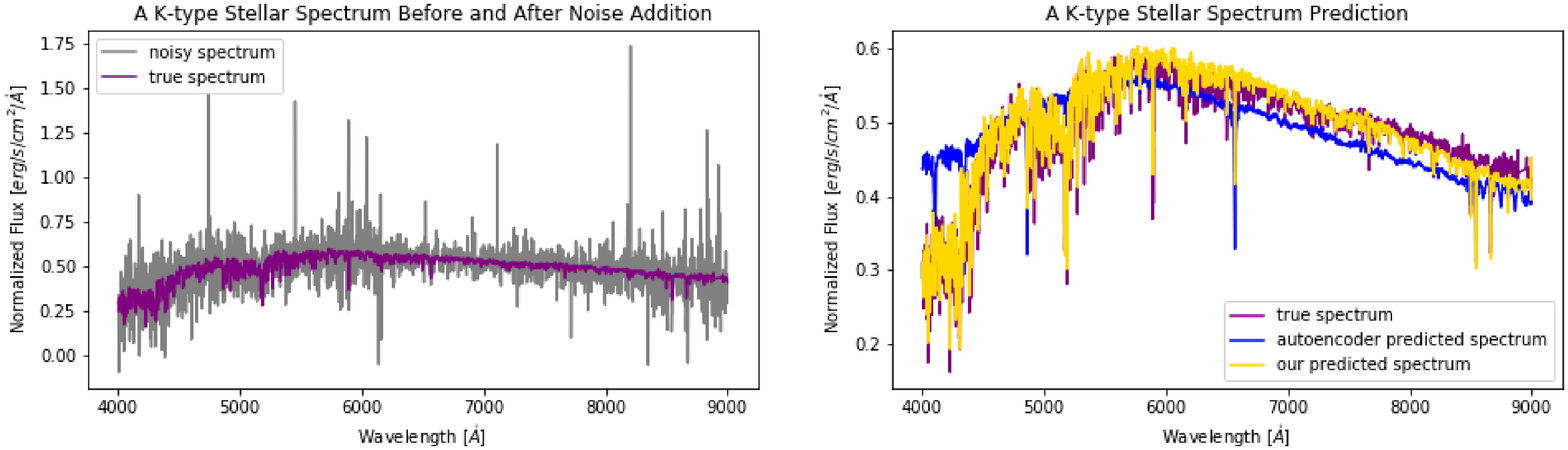}
	\includegraphics[scale=0.4]{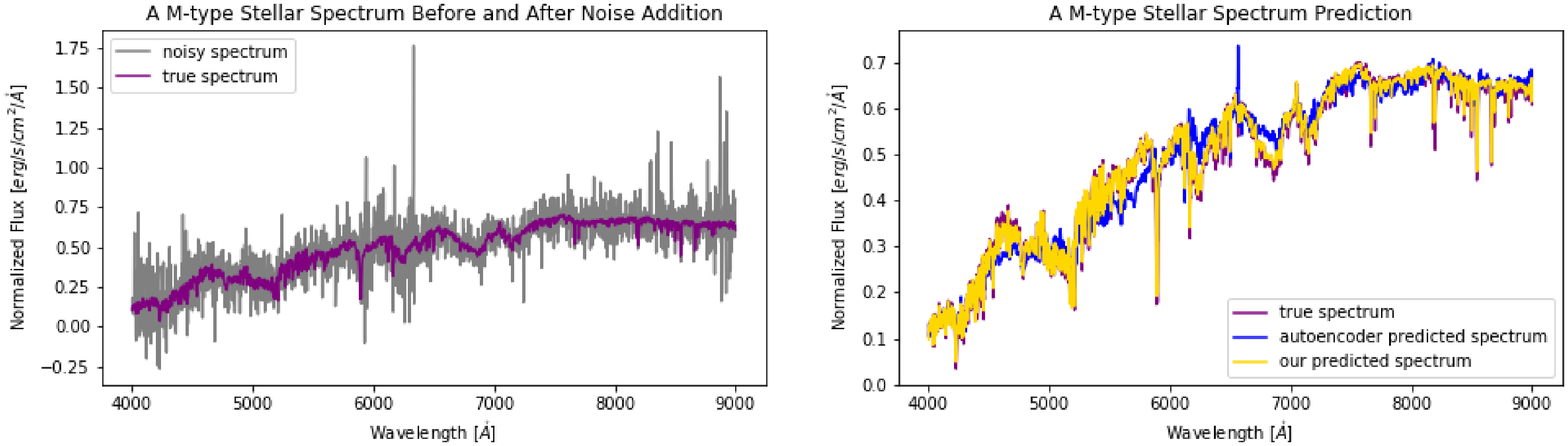}
	\includegraphics[scale=0.4]{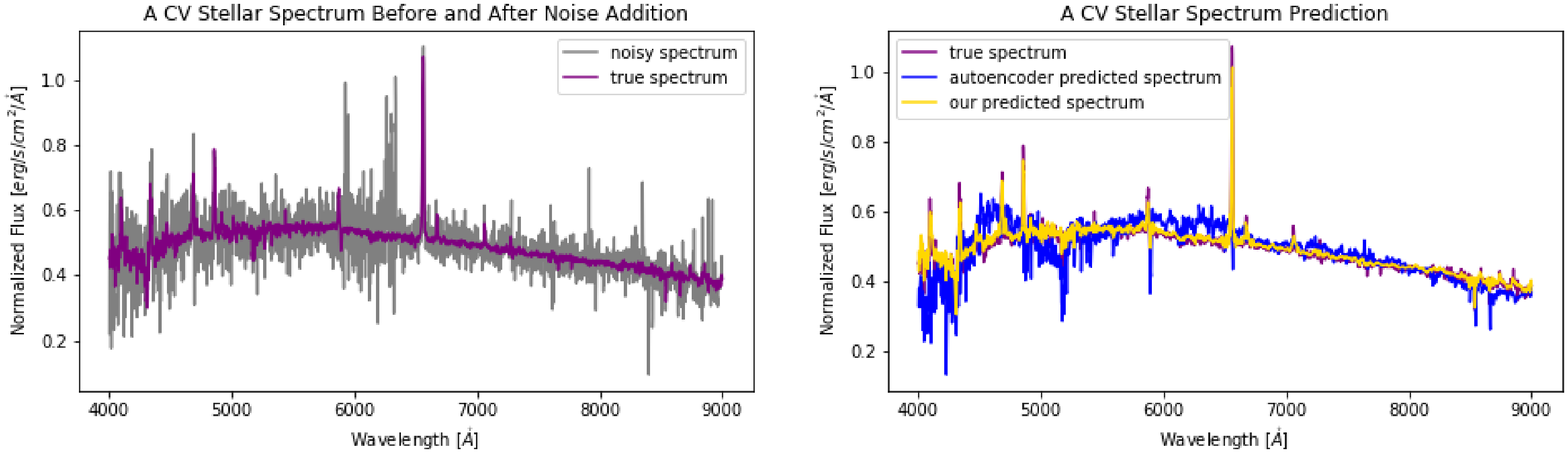}
	\includegraphics[scale=0.4]{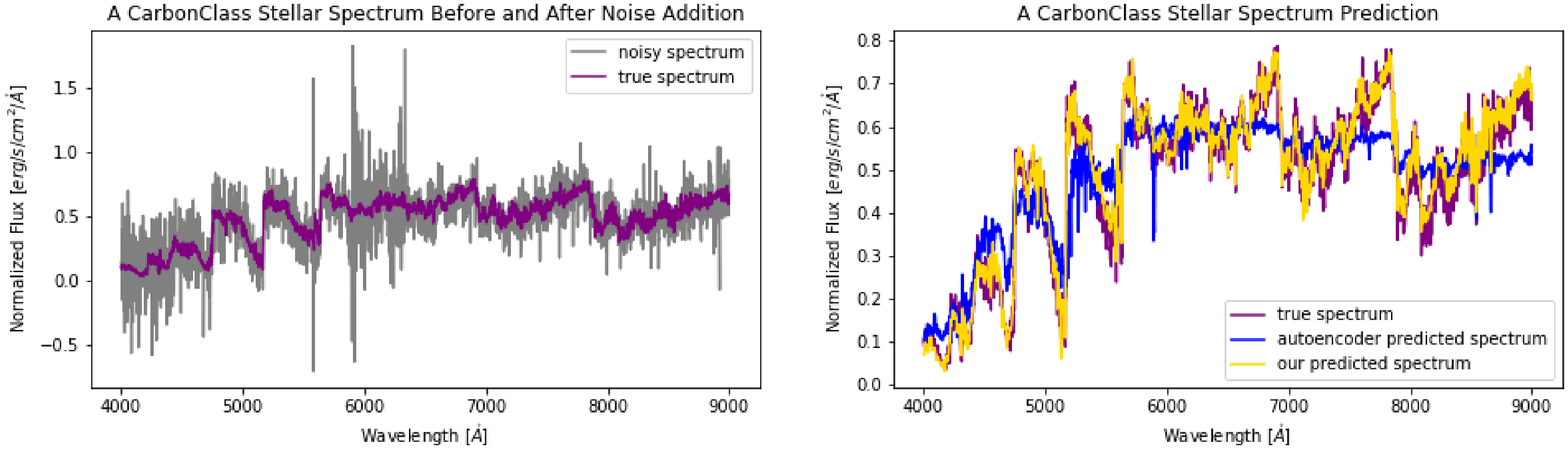}	
	\includegraphics[scale=0.4]{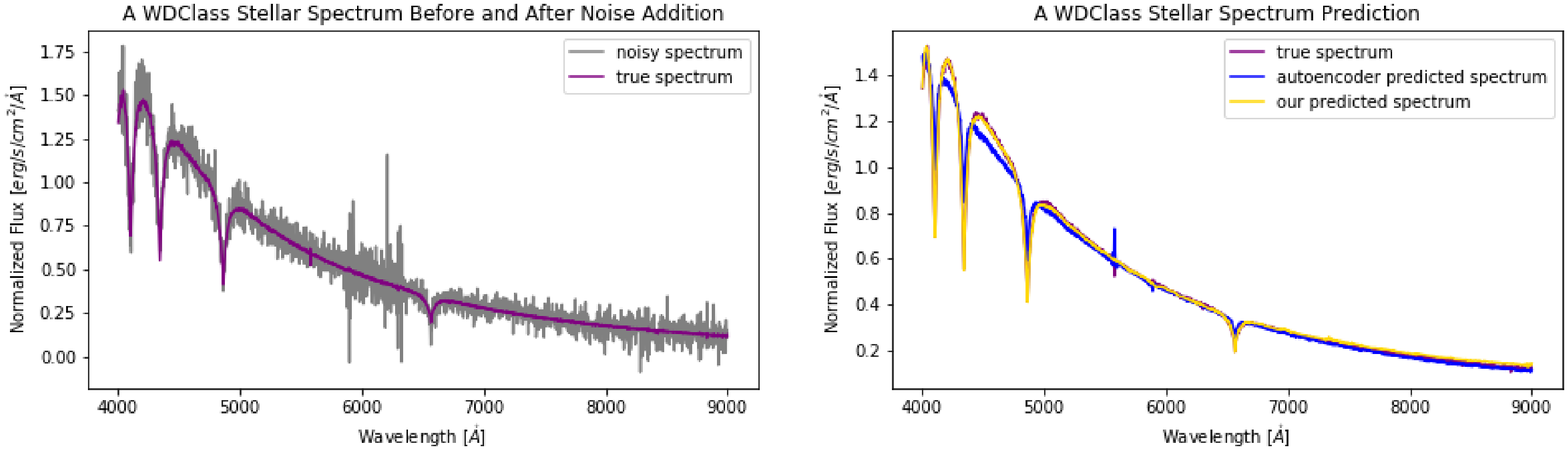}
	\caption{The five rows show examples of stellar spectra denoised by the convolutional denoising auto-encoder and our method for the stellar subclasses K, M, CV, Carbon, WD, respectively. The left column shows the true spectrum and its synthetic counterpart with noise. The purple curve is the clean spectrum that both denoising algorithms try to recover, and the grey curve is the  spectrum added with  noise. The  spectra denoised by both algorithms are compared  in the right column. The denoised spectrum from our model (yellow curve) is much closer to the purple clean spectrum than the standard auto-encoder result (blue curve).
	\label{fig:testing2}}
\end{figure}

Figure~\ref{fig:testing1} and Figure~\ref{fig:testing2} exhibit some examples of denoised spectra for ten stellar subclasses. We choose one representative result from each of the ten stellar subclasses to demonstrate the power of our method. The left column shows the spectrum before and after noise contamination, where the purple curve is the high-SNR spectrum and the grey curve is the clean spectrum with added noise. The noisy spectrum gets cleaned by the standard auto-encoder and our proposed method. The denoised spectrum is shown in the right column of Figure~\ref{fig:testing1} and Figure~\ref{fig:testing2}. In each subfigure of the right column, the purple spectrum is the true spectrum that both denoising algorithms try to recover. Though the clean purple spectrum is unknown to the denoising algorithm, our proposed method shows promising ability to recover it from the noisy observation. The predicted spectrum from our model (yellow curve) is much closer to the purple clean spectrum than the standard auto-encoder result (blue curve). Our proposed method also has the capacity to remove strong noisy emission lines (see the second row of Figure~\ref{fig:testing2}) and keeps the signal emission lines (see the third row of Figure~\ref{fig:testing2}).

Figure~\ref{fig:lossDenoise} shows the overall comparison between our method and the convolutional denoising auto-encoder in spectrum denoising for various stellar subclasses. Each yellow point in a subfigure represents one synthetic noisy stellar spectrum. The noisy spectrum gets cleaned and compared with the true high-SNR spectrum, and the reconstruction loss is computed. The reconstruction loss is computed as follows. Suppose $\vvs$ is the true signal spectrum and $\hat{\vvs}$ is the denoised spectrum by a model. Then, the reconstruction loss is $\sum_{d=1}^{D} (s_d - \hat{s}_d)^2/D$.
In each subfigure, the horizontal axis is the reconstruction loss for the convolutional denoising auto-encoder, and the vertical axis is the reconstruction loss of our method. The blue line indicates where the two methods have equal performance. We can see that most points in each subfigure fall below the blue line, indicating that our method has smaller reconstruction loss. The title of each subfigure reports the proportion of spectra for which our method has smaller reconstruction loss. Within each stellar subclass, our method produces higher-quality denoised spectra for more than 80\% of the testing samples. For subclasses such as Carbon class and WD class, our method demonstrates improved performance for almost 100\% of the test samples.

\begin{figure}
	\centering
	\includegraphics[scale=0.45]{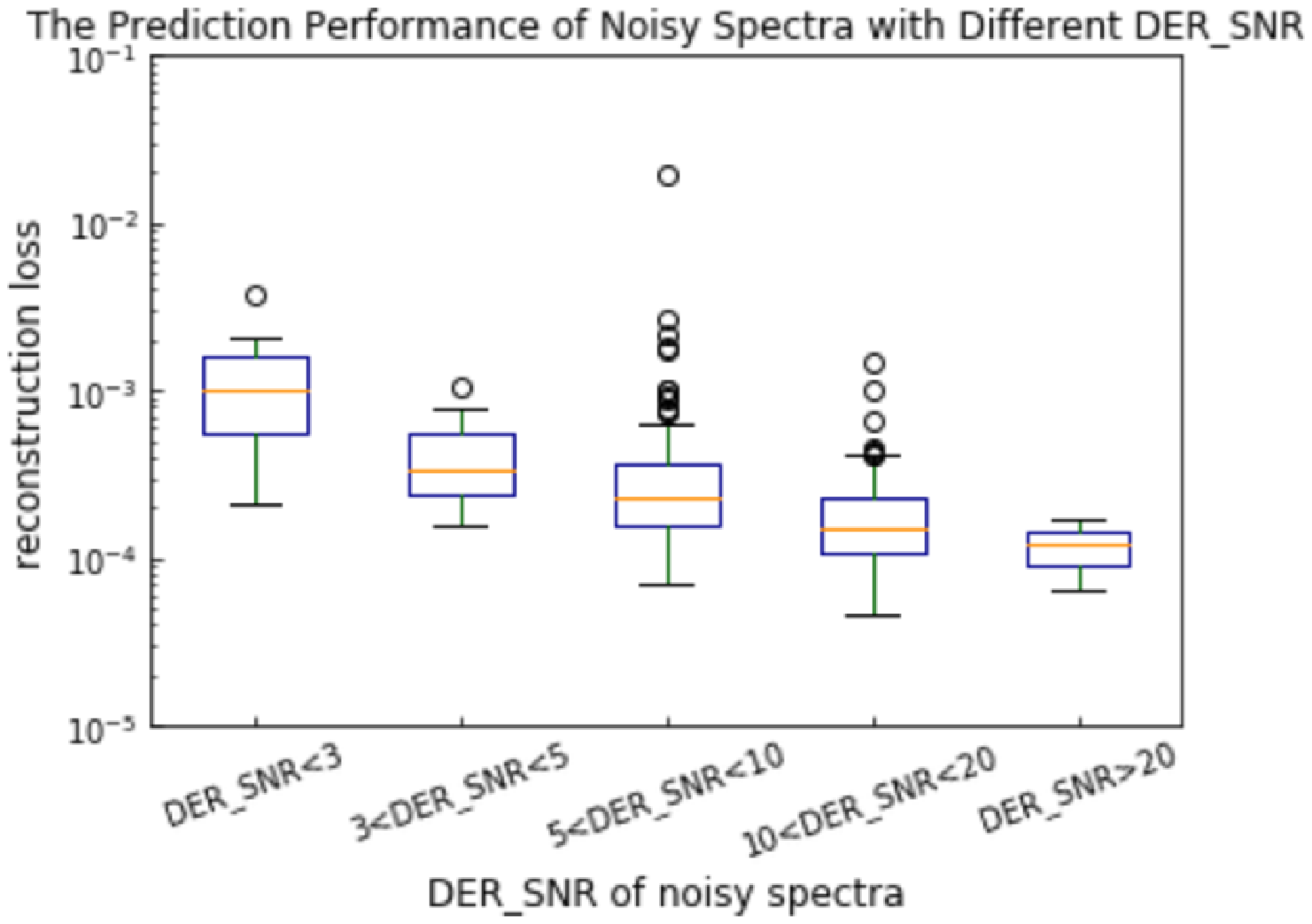}
		\includegraphics[scale=0.45]{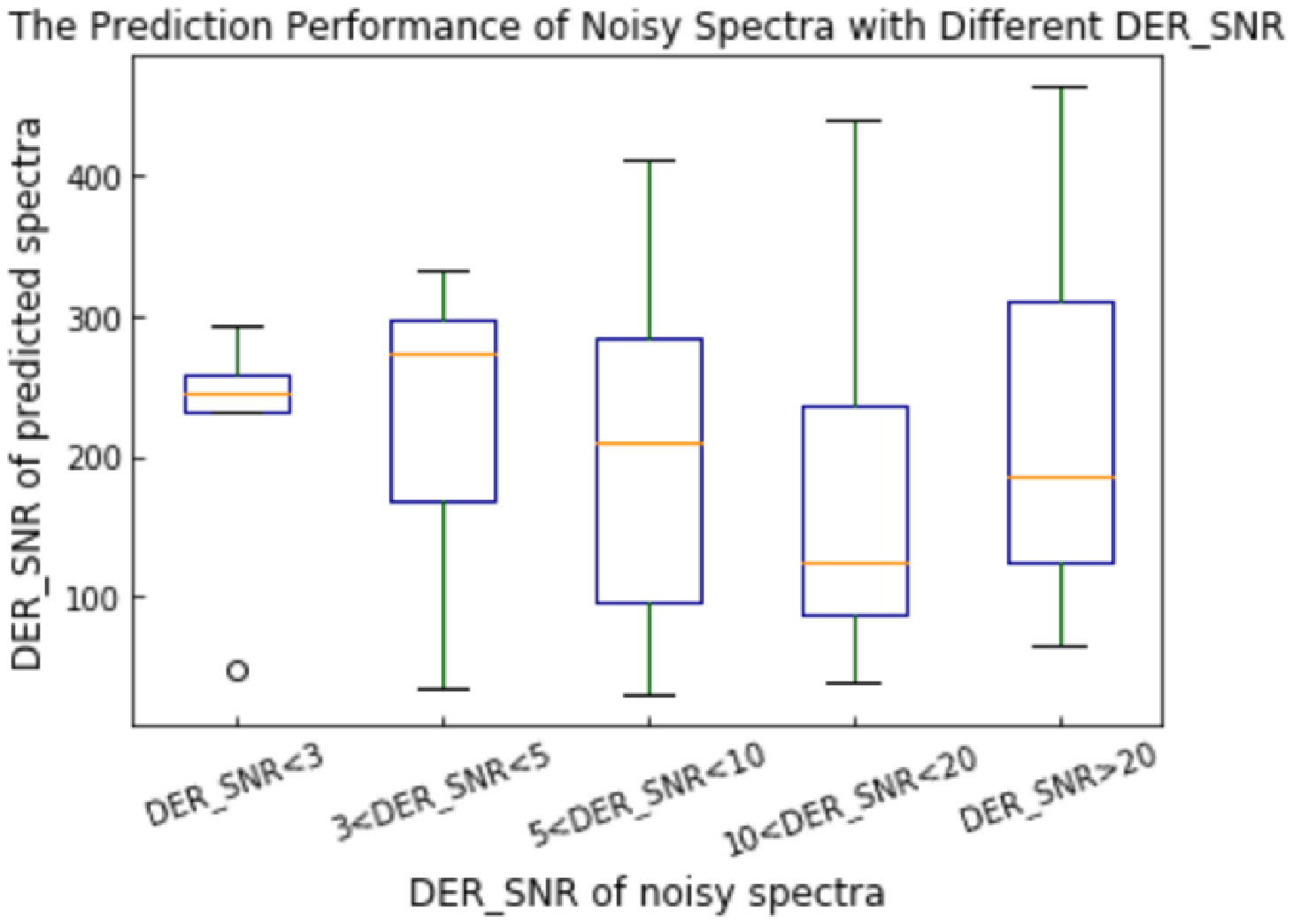}
	\caption{Model performance under varying levels of signal-to-noise ratio (DER\_SNR). The left panel shows how the reconstruction loss (vertical axis) depends on the DER\_SNR of the input synthetic spectrum (horizontal axis). The right panel shows how the DER\_SNR of the denoised spectrum (vertical axis) varies with the DER\_SNR of the input synthetic spectrum (horizontal axis). The reconstruction loss does not increase very quickly as DER\_SNR decreases in the left panel, while the  DER\_SNR of the denoised spectra is stable in the right panel. }
	\label{fig:snr}
\end{figure}

\begin{figure}
        \includegraphics[scale=0.36]{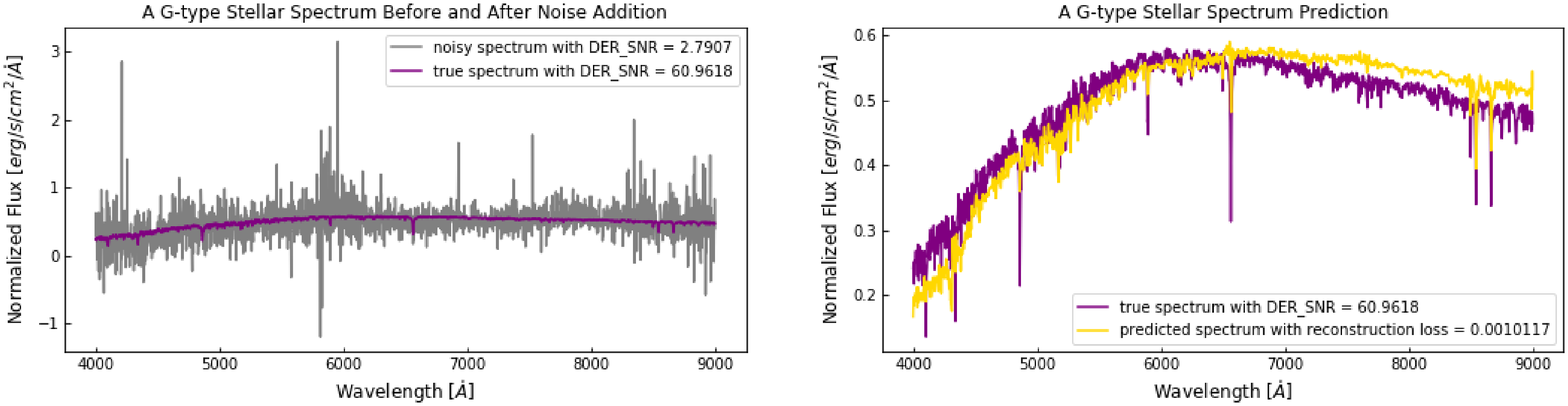}
        \caption{An illustrative example. In the left panel, the DER\_SNR of the synthetic noisy spectrum (grey curve) is 2.79 and the DER\_SNR of the true spectrum (purple curve) is 60.96. In the right panel, the reconstruction loss between the true spectrum (purple curve) and the predicted spectrum from our model (yellow curve) is $1.0\times 10^{-3}$. } \label{fig:snr:eg1}
\end{figure}

\begin{figure}
	\includegraphics[scale=0.36]{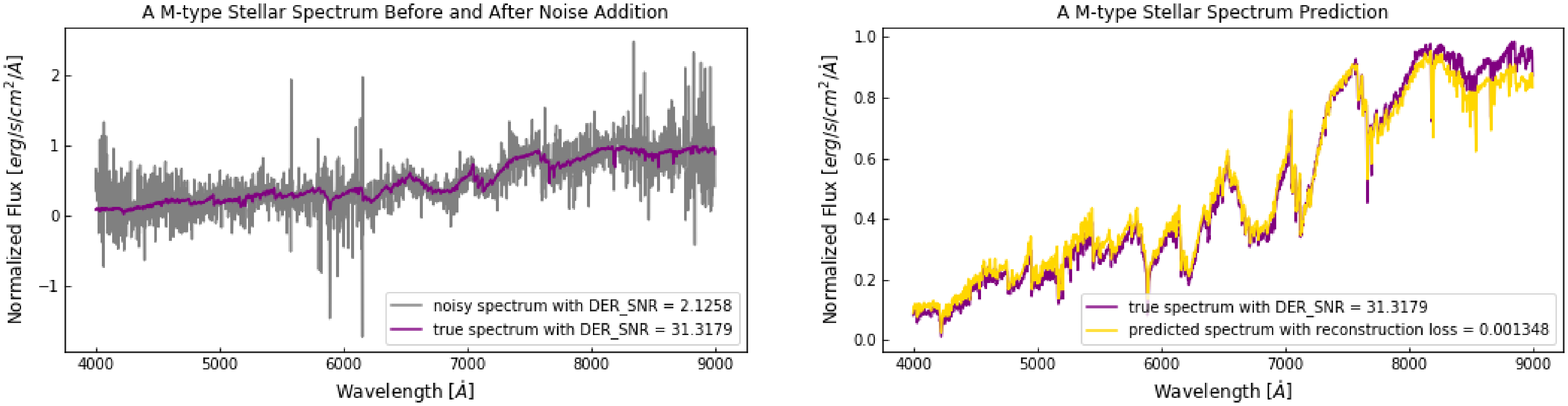}
	  \caption{An illustrative example. In the left panel, the DER\_SNR of the synthetic noisy spectrum (grey curve) is 2.13 and the DER\_SNR of the true spectrum (purple curve) is 31.32. In the right panel, the reconstruction loss between the true spectrum (purple curve) and the predicted spectrum from our model (yellow curve) is $1.3\times 10^{-3}$.} \label{fig:snr:eg2}
\end{figure}

We further consider how the signal-to-noise ratio affects our model performance. For the synthetic spectra, we measure their signal-to-noise ratio by the formula given by \cite{stoehr2008der_snr}. The computed signal-to-noise ratio is denoted as DER\_SNR. To decrease DER\_SNR of the synthetic data to a specific level, we add additional gaussian noise with various noise levels to each spectrum. This procedure results in a new group of test dataset. The left panel of Figure~\ref{fig:snr} shows how boxplot of the reconstruction loss varies across distinct DER\_SNR levels. The reconstruction loss does not increase very quickly as DER\_SNR decreases. When the  DER\_SNR is below 3, the median reconstruction loss is still about $10^{-3}$. To illustrate how the spectrum looks like at this level of reconstruction loss, we plot a few examples in Figures~\ref{fig:snr:eg1}--\ref{fig:snr:eg2}. The grey curve in the left panel of Figure~\ref{fig:snr:eg1} is one synthetic spectrum with DER\_SNR equaling to $2.79$. The reconstruction loss between the true spectrum and the denoised spectrum in the right panel is about $1.0\times 10^{-3}$. Figure~\ref{fig:snr:eg2} shows one more example where the DER\_SNR of the synthetic spectrum is $2.12$ and the reconstruction loss is about $1.3\times 10^{-3}$. The right panel of Figure~\ref{fig:snr} compares the DER\_SNR before and after denoising for the above dataset. For most spectra, their DER\_SNR after denoising is above 100. From the right panel of Figure~\ref{fig:snr}, we can also find that the DER\_SNR of our model output is stable regardless of the DER\_SNR of the input spectrum. This is because our model prior component and the generator in \eqref{eqn:model:level12}-- \eqref{eqn:model:level3} are fixed after model training. The denoised spectra, which are generated by the prior and the generator,  will always have the same level of SNR of the training dataset.

\subsection{Spectrum Denoising with Missing Flux}
\label{sec:test:noise2}

Besides the above test dataset, another test dataset is created to evaluate model performance for spectrum denoising with missing flux values. To construct this benchmark data, almost the same procedure of Section~\ref{sec:test:noise1} is taken. Realistic noise is added to the clean spectrum with high SNR. In addition, we randomly remove the flux values over an interval range of wavelength. The interval of missing pixels is also randomly selected for each test spectrum.

Based on the idea of masked likelihood~\eqref{eqn:maskedlikelihood}, our trained model can be directly employed to denoise spectrum with partially missing flux values. In other words, our model does not require re-training to deal with this kind of data. However, the standard denoising convolutional auto-encoder requires re-training to adapt to this dataset. Its original training data also gets randomly removed flux values over a random wavelength range. The missing values are imputed by zeros, and the denoising convolutional auto-encoder aims to reconstruct the original full spectrum from the zero-imputed spectrum. The full spectrum is employed for its loss computation and parameter update.

\begin{figure}[t]
	\centering
	\includegraphics[scale=0.32]{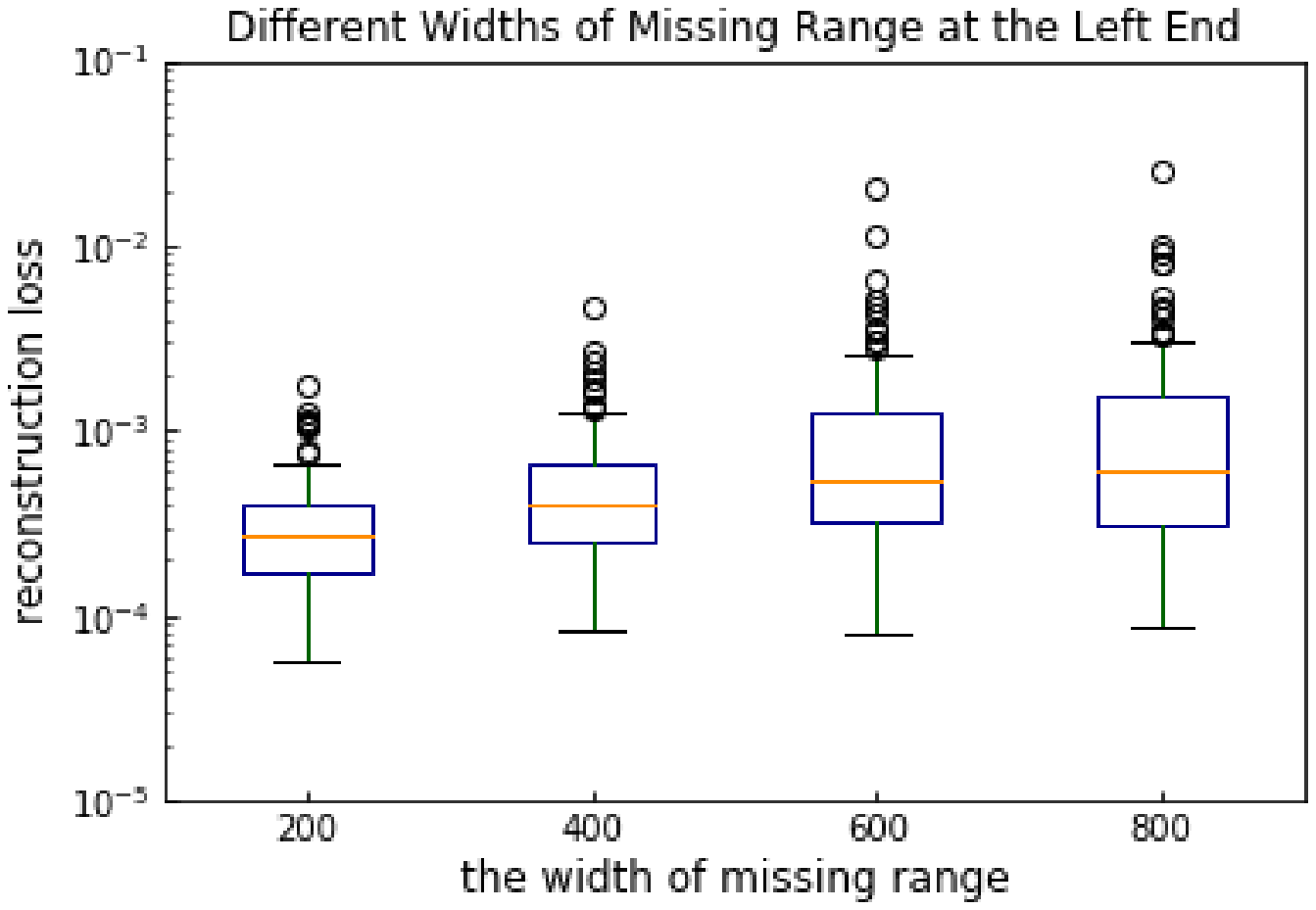}~\includegraphics[scale=0.32]{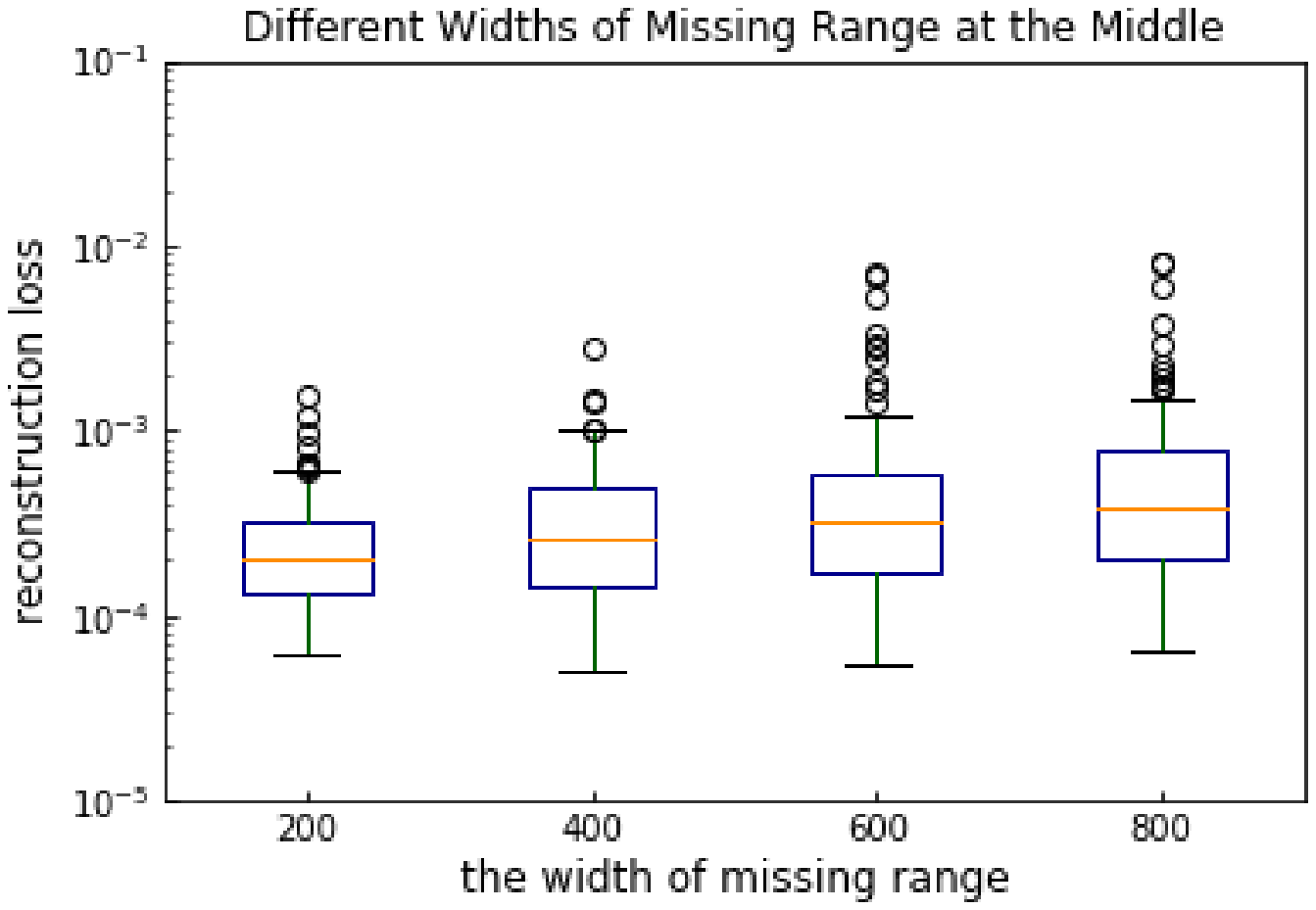}~\includegraphics[scale=0.32]{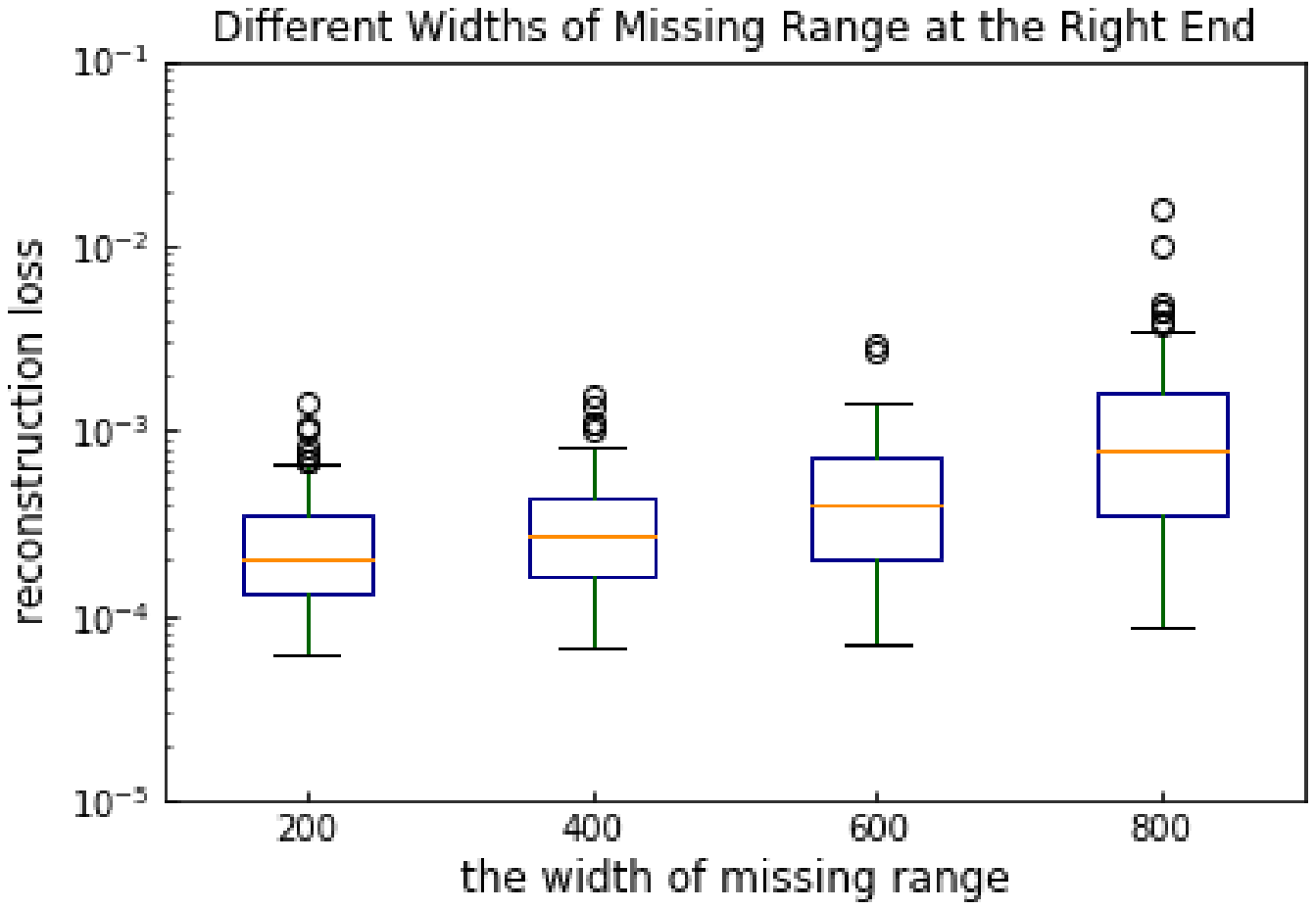}
	\caption{The three panels correspond to the cases where the missing flux values occur at the  left end, middle and right end of each spectrum, respectively. The horizontal axis represents the  width of the missing range. The vertical axis is the reconstruction loss of our model. The reconstruction loss increases with the growing number of missing pixels. The model performance is not sensitive to the location of the missing pixels, but missing at the left end (blue end) incurs slightly higher loss as the blue end is feature-rich for most spectra.
	 \label{fig:missing_position_width}}
\end{figure}

Figure~\ref{fig:missing_position_width} shows the performance of our model depending on different positions and widths of the missing range. The three panels correspond to the cases where the missing flux occurs at the left end (blue end), the middle and the right end (red end) of the spectrum, respectively. In each panel, the horizontal axis is the number of missing pixels out of the total $D=2048$ pixels. The vertical axis is the reconstruction loss. As expected, the reconstruction loss increases with the growing number of missing pixels. When the length of missing pixels is 800 (i.e. 40\% of the whole spectrum), the median reconstruction loss is still below $10^{-3}$. Generally, the model performance is  not sensitive to the location of the missing pixels, but missing at the left end (blue end) incurs slightly higher loss. This is due to the fact that the blue end is feature-rich and contains more information for spectrum reconstruction for most spectra.

\begin{figure}
	\centering
    \begin{minipage}{0.49\linewidth}
        \includegraphics[scale=0.42]{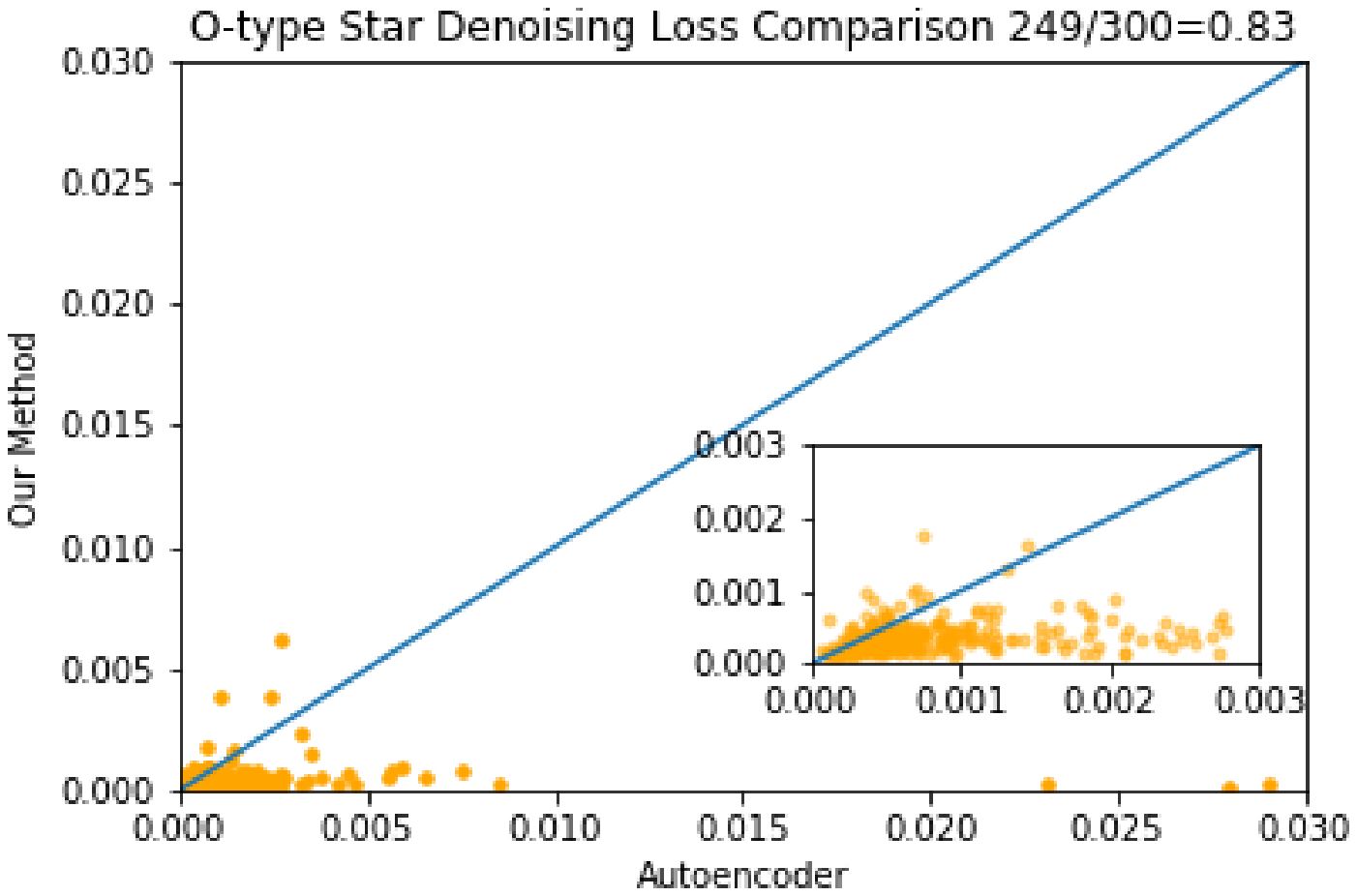}
    \end{minipage}
    \begin{minipage}{0.49\linewidth}
        \includegraphics[scale=0.42]{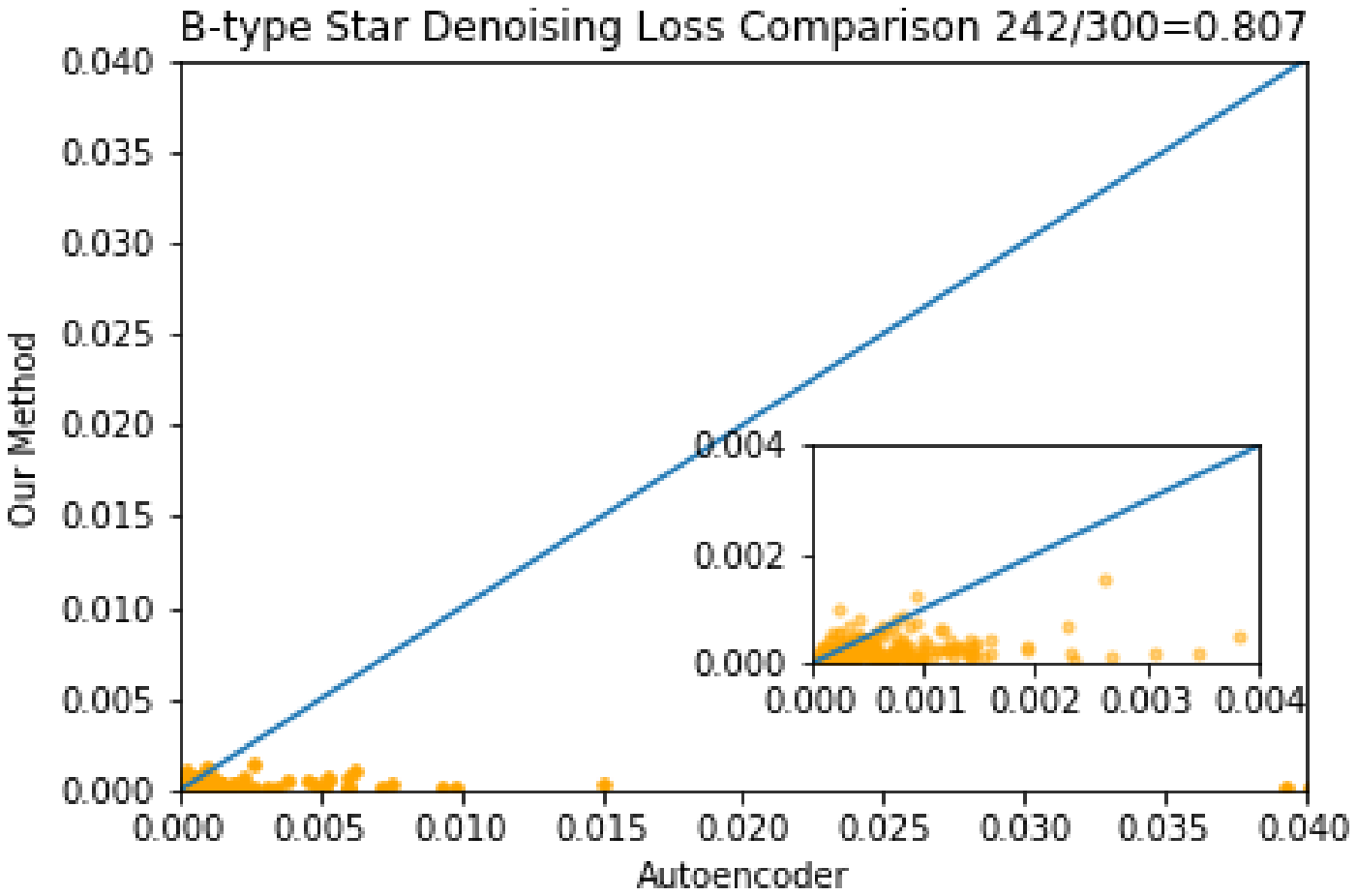}
    \end{minipage}
    \begin{minipage}{0.49\linewidth}
        \includegraphics[scale=0.42]{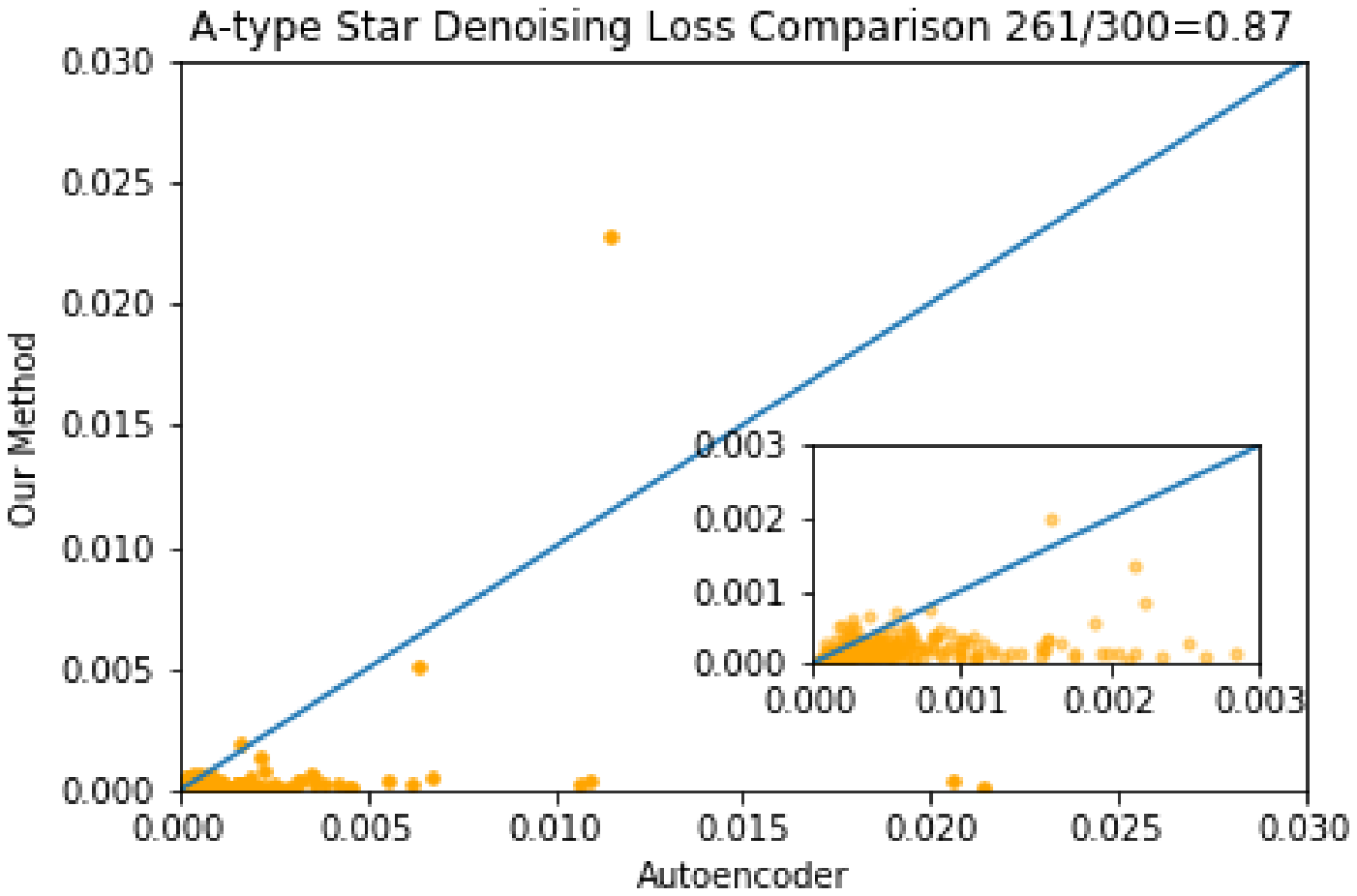}
    \end{minipage}
    \begin{minipage}{0.49\linewidth}
        \includegraphics[scale=0.42]{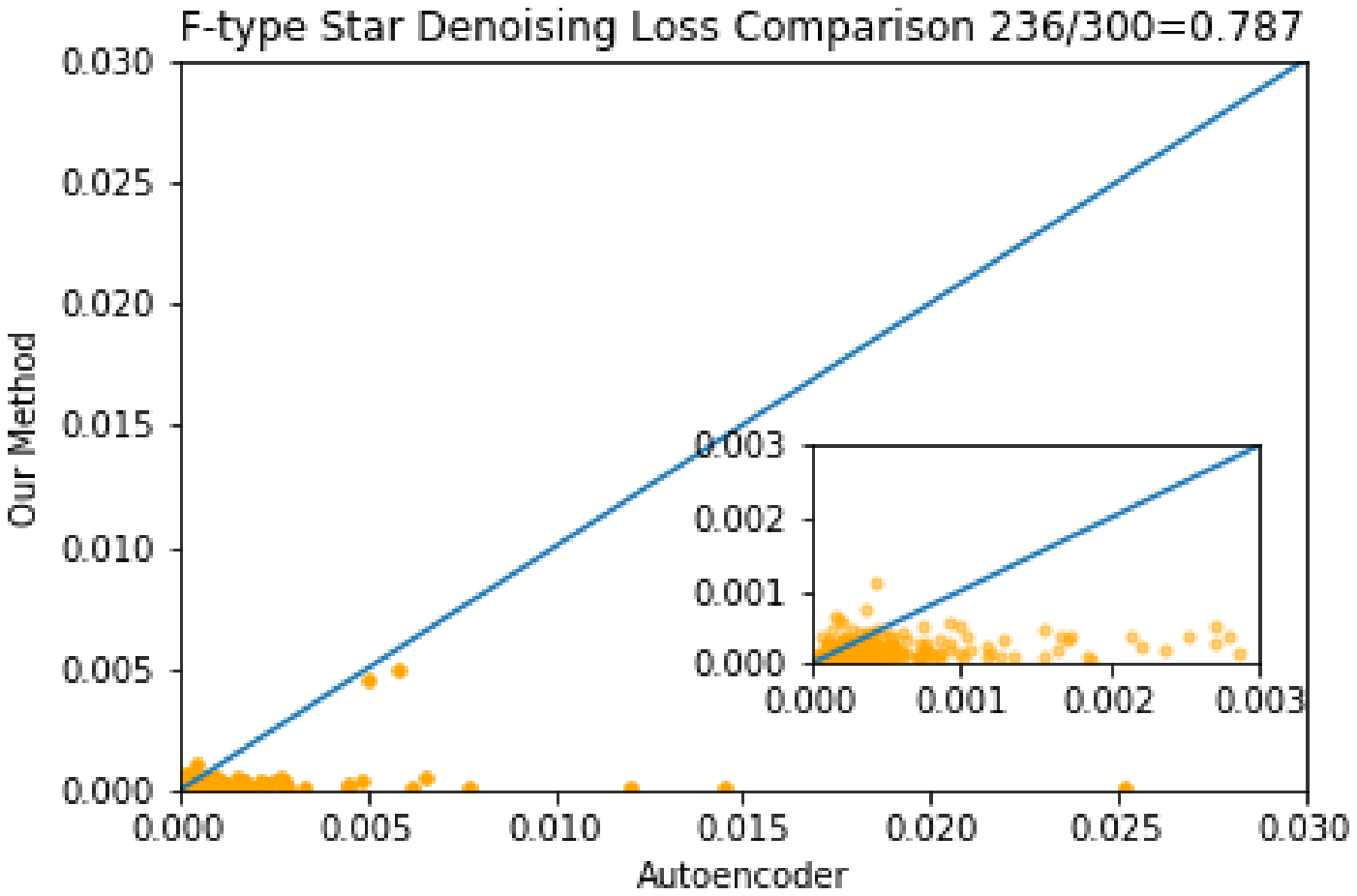}
    \end{minipage}
     \begin{minipage}{0.49\linewidth}
        \includegraphics[scale=0.42]{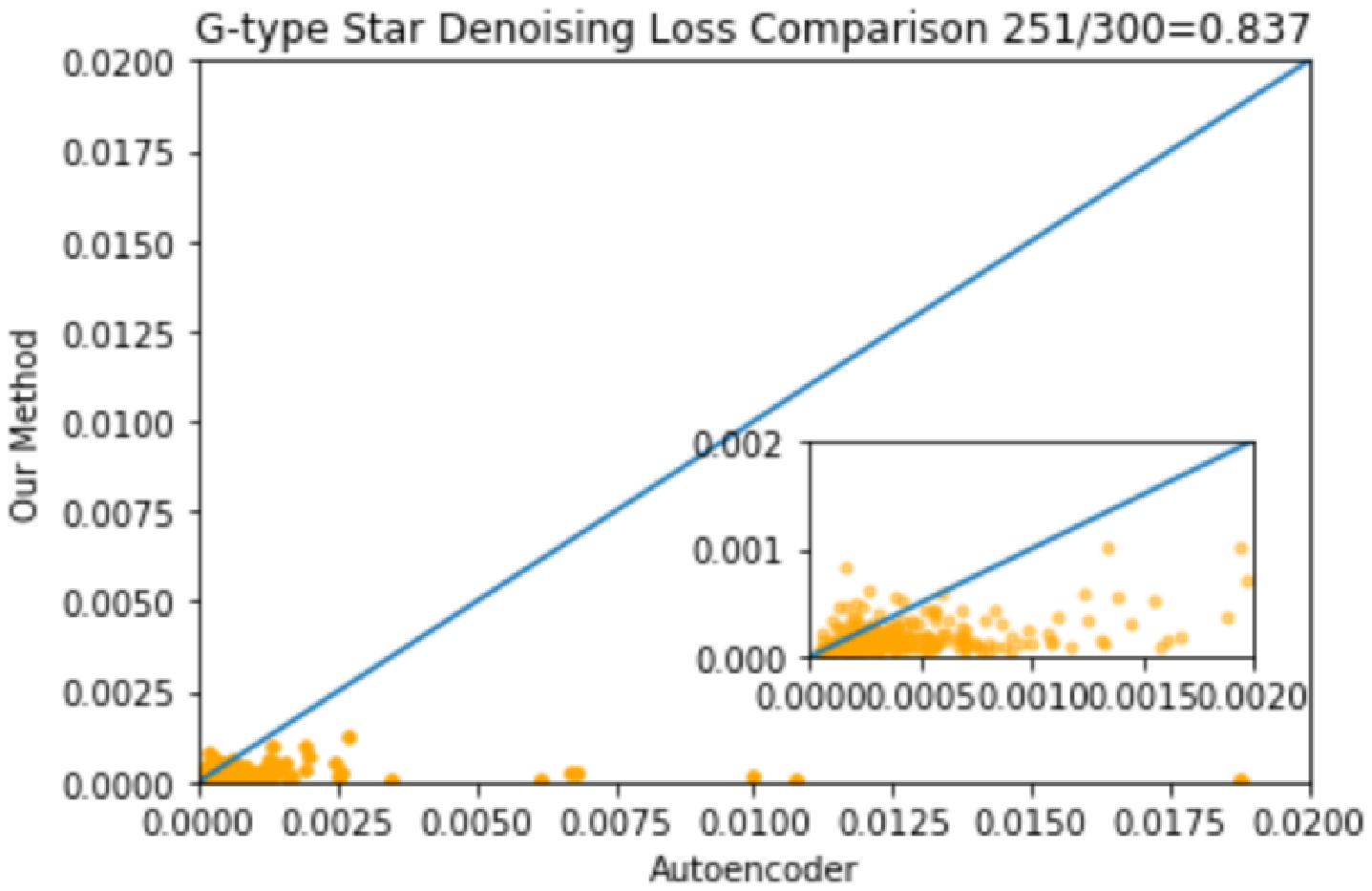}
    \end{minipage}
    \begin{minipage}{0.49\linewidth}
        \includegraphics[scale=0.42]{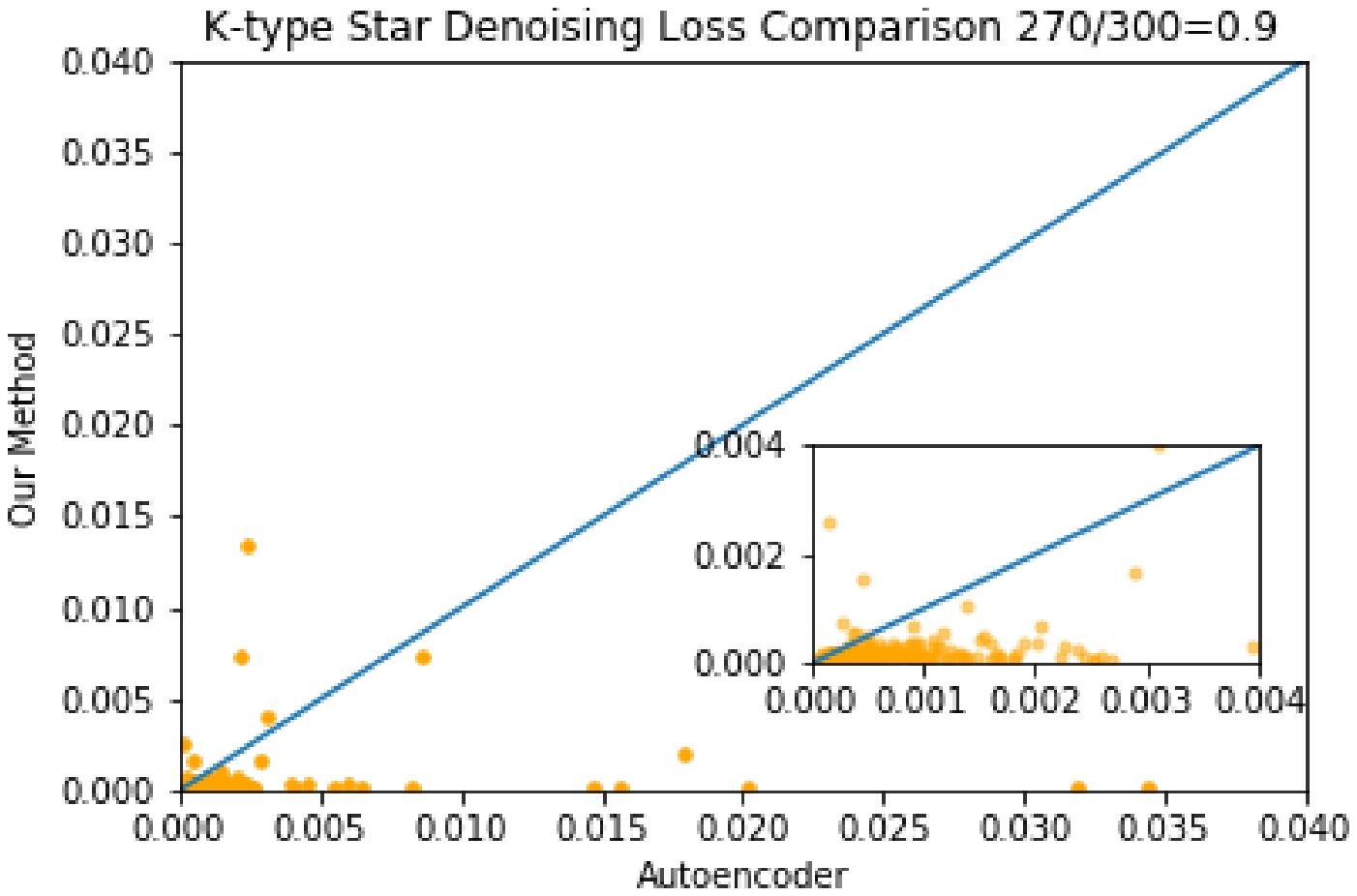}
    \end{minipage}
     \begin{minipage}{0.49\linewidth}
        \includegraphics[scale=0.42]{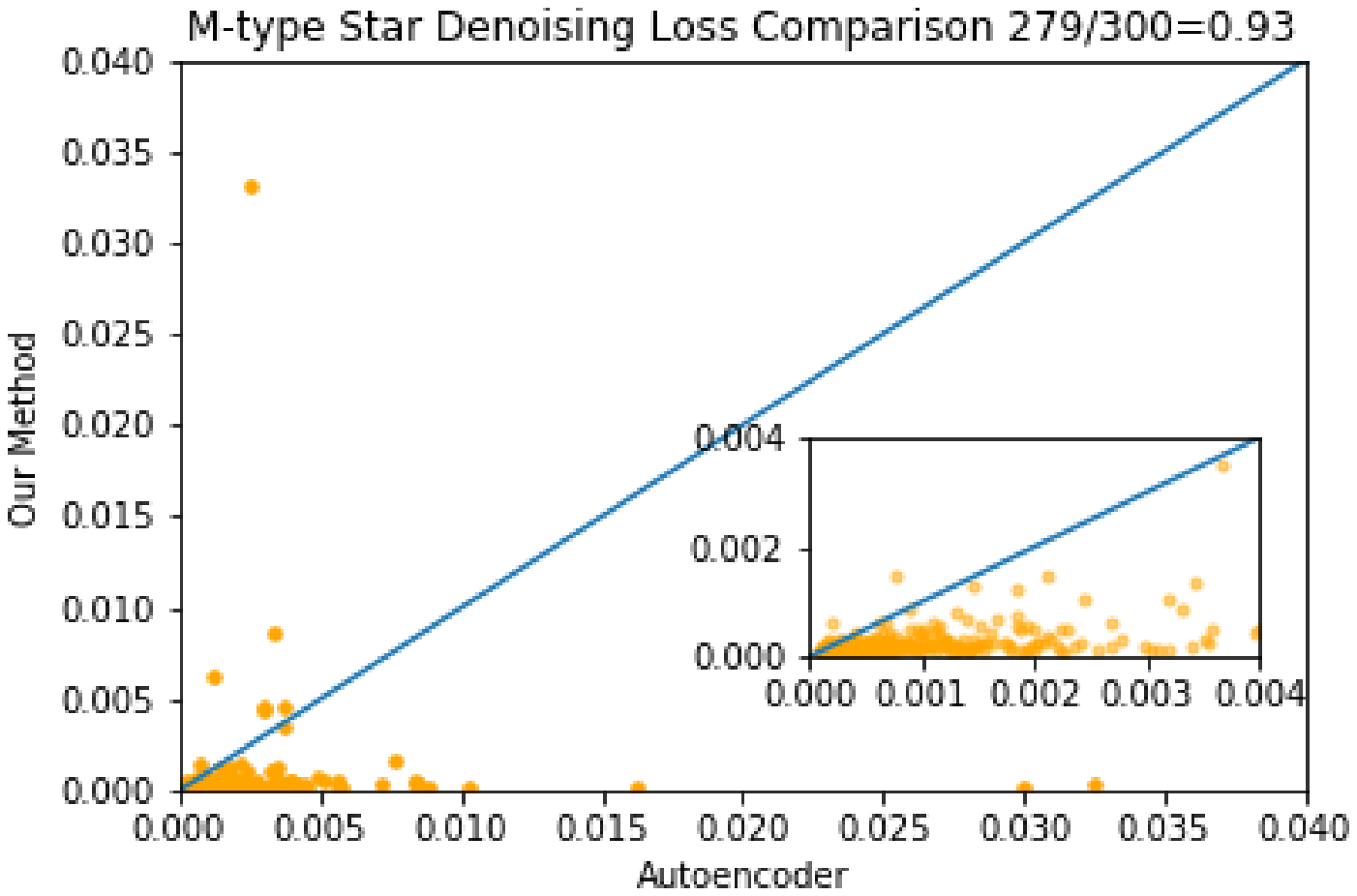}
    \end{minipage}
    \begin{minipage}{0.49\linewidth}
        \includegraphics[scale=0.42]{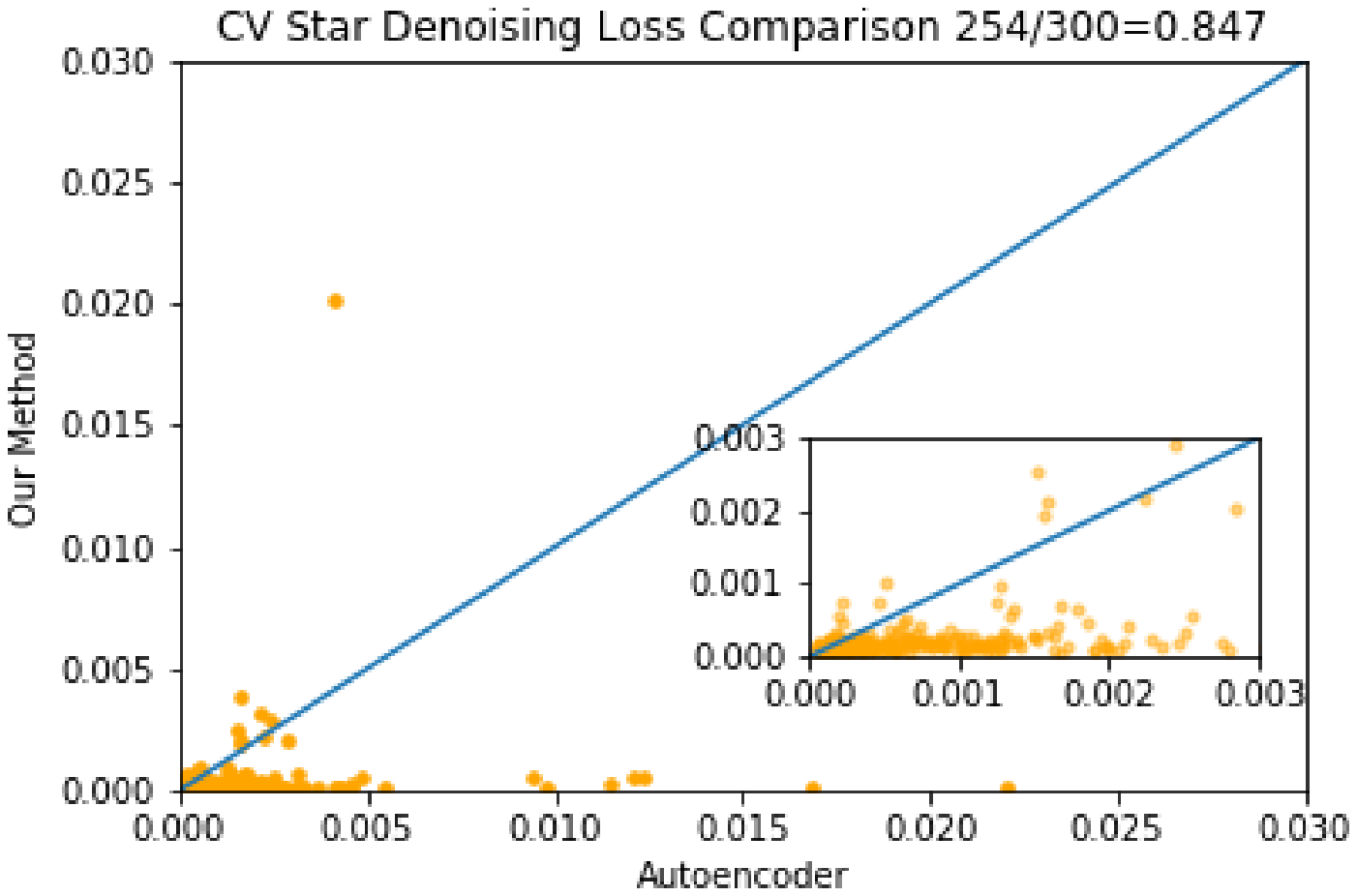}
    \end{minipage}
     \begin{minipage}{0.49\linewidth}
        \includegraphics[scale=0.42]{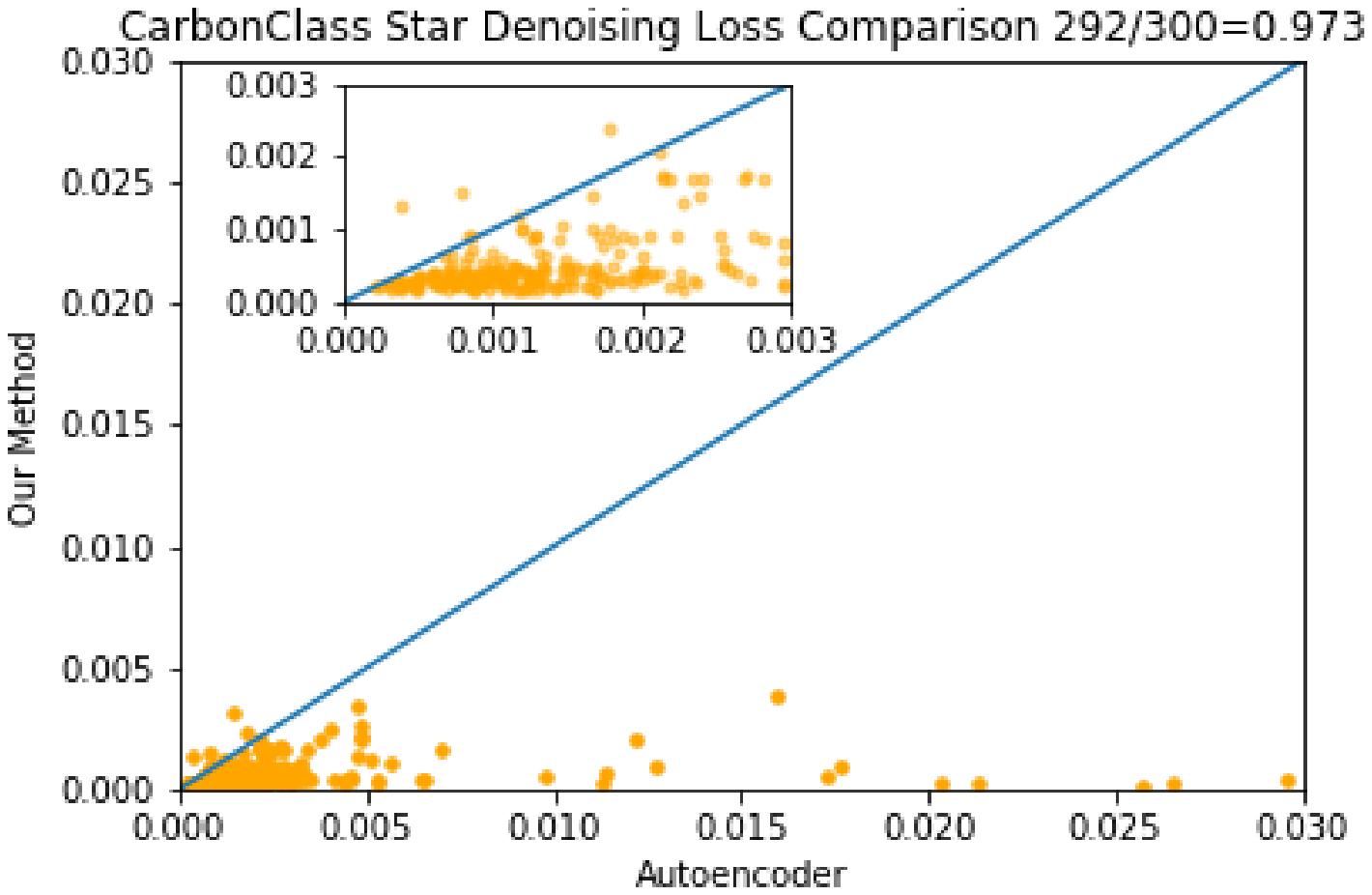}
    \end{minipage}
    \begin{minipage}{0.49\linewidth}
        \includegraphics[scale=0.42]{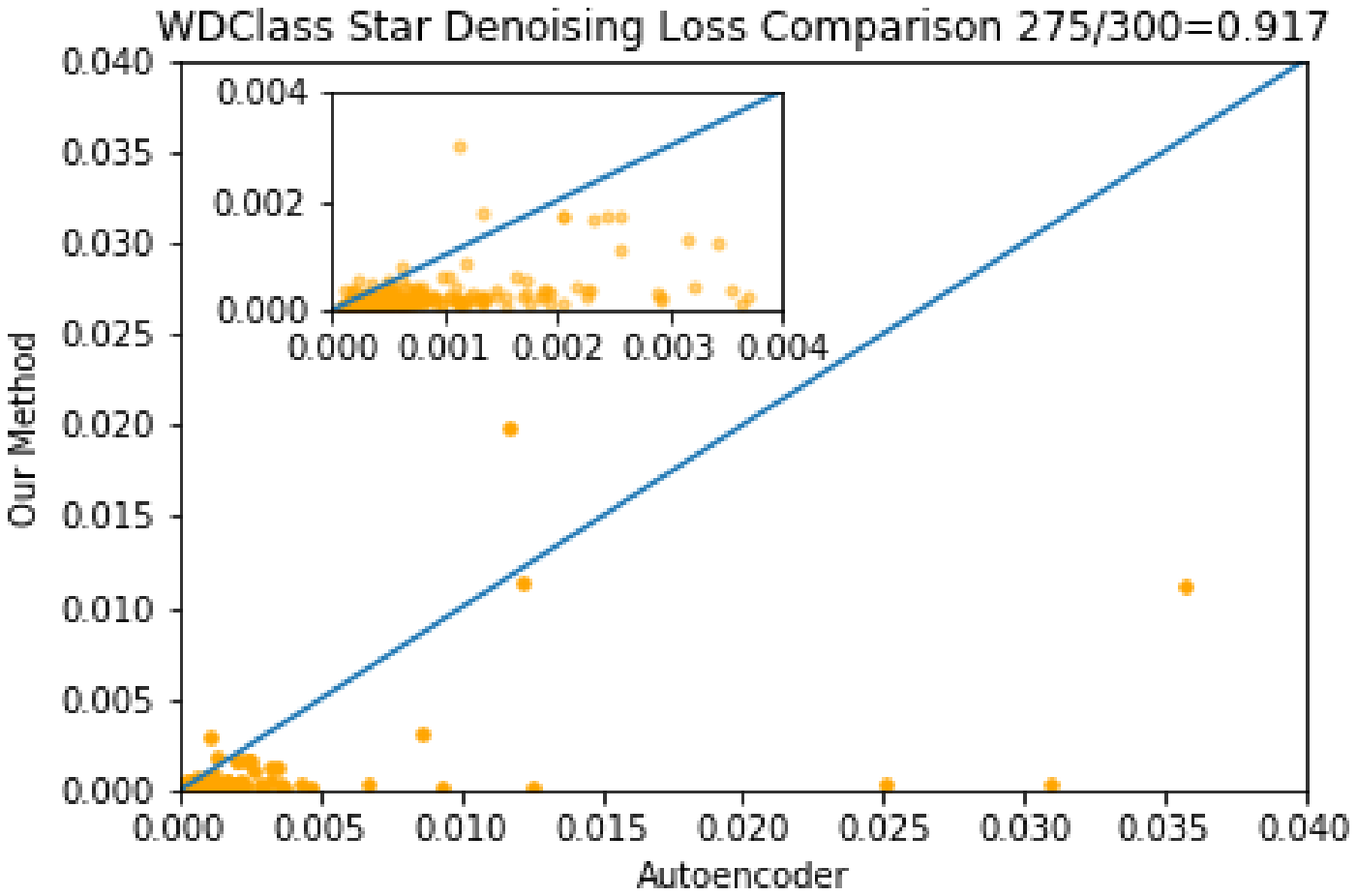}
    \end{minipage}
	\caption{The loss comparison for the convolutional auto-encoder and our method for the ten stellar subclasses. The horizontal axis is the reconstruction loss for the convolutional denoising auto-encoder, and the vertical axis is the reconstruction loss for our method. Each yellow point in a subfigure corresponds to one synthetic noisy spectrum. The blue line indicates where the two methods have equal performance. Most points in each subfigure fall below the blue line, indicating that our method has smaller reconstruction loss. The title of each subfigure also reports the proportion of spectra for which our method has smaller reconstruction loss. Our method demonstrates improved performance.}
	\label{fig:lossDenoise}
\end{figure}

\begin{figure}
	\centering
	\includegraphics[scale=0.4]{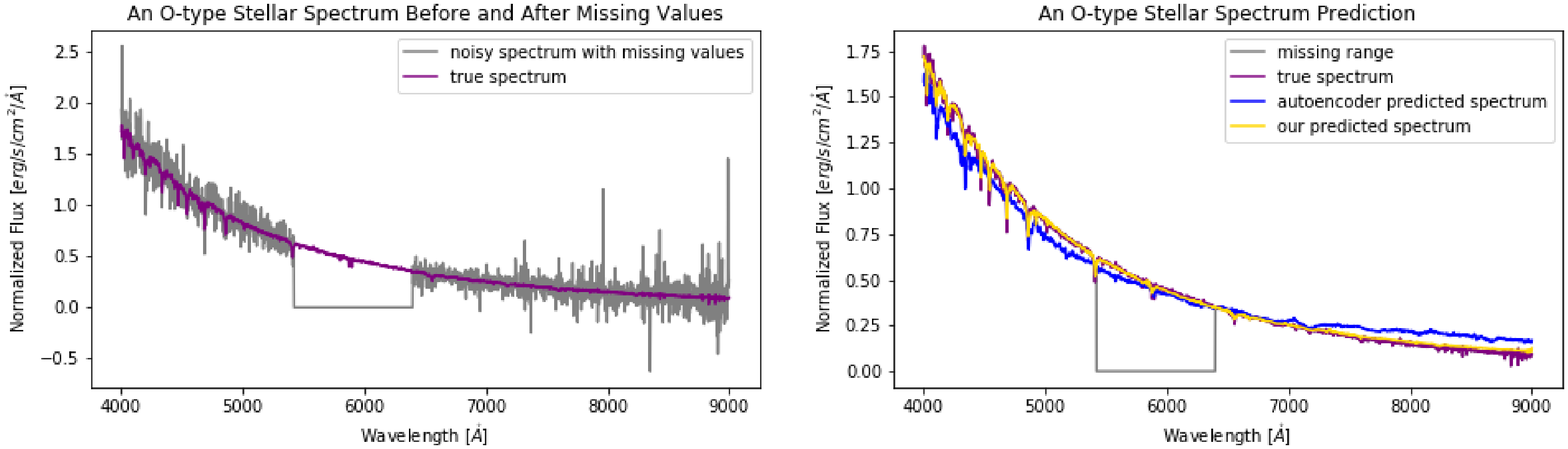}
	\includegraphics[scale=0.4]{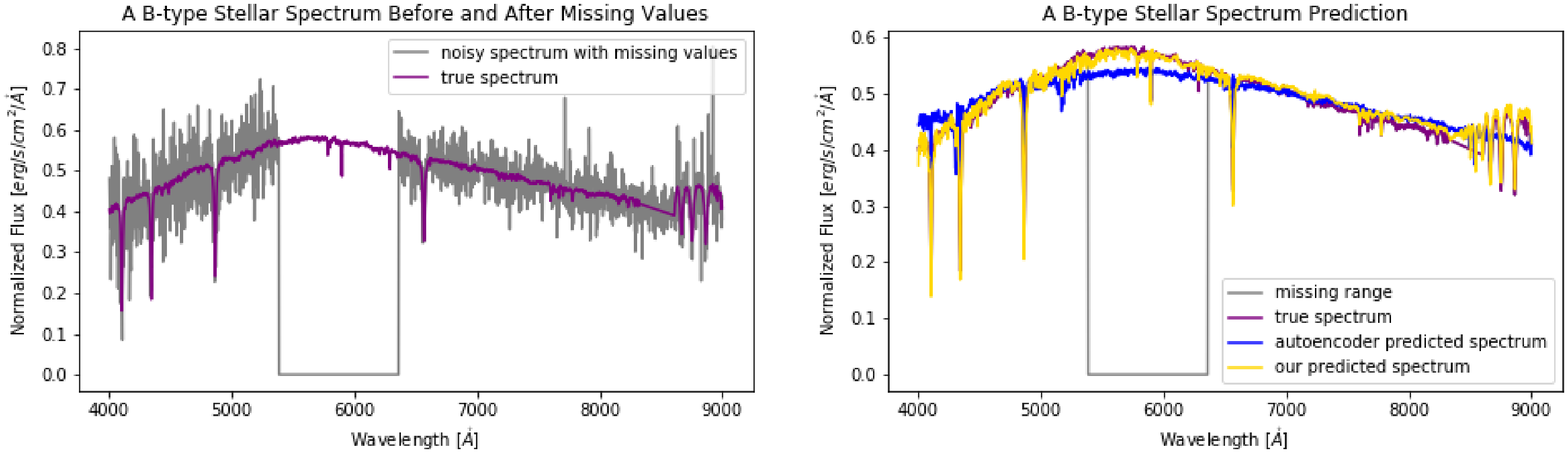}
	\includegraphics[scale=0.4]{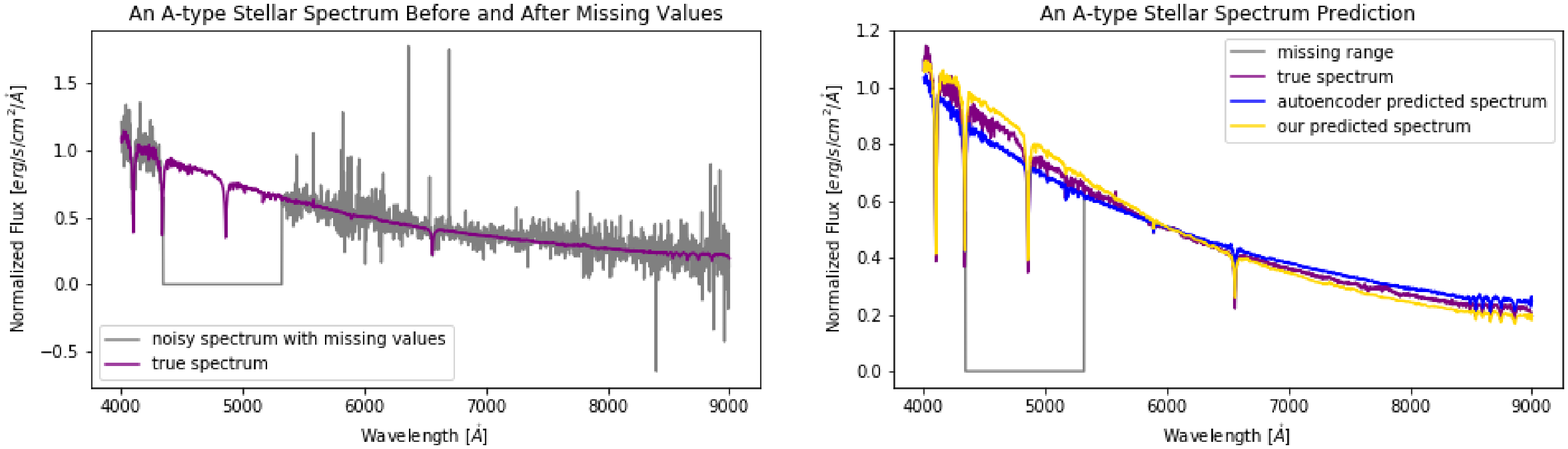}
	\includegraphics[scale=0.4]{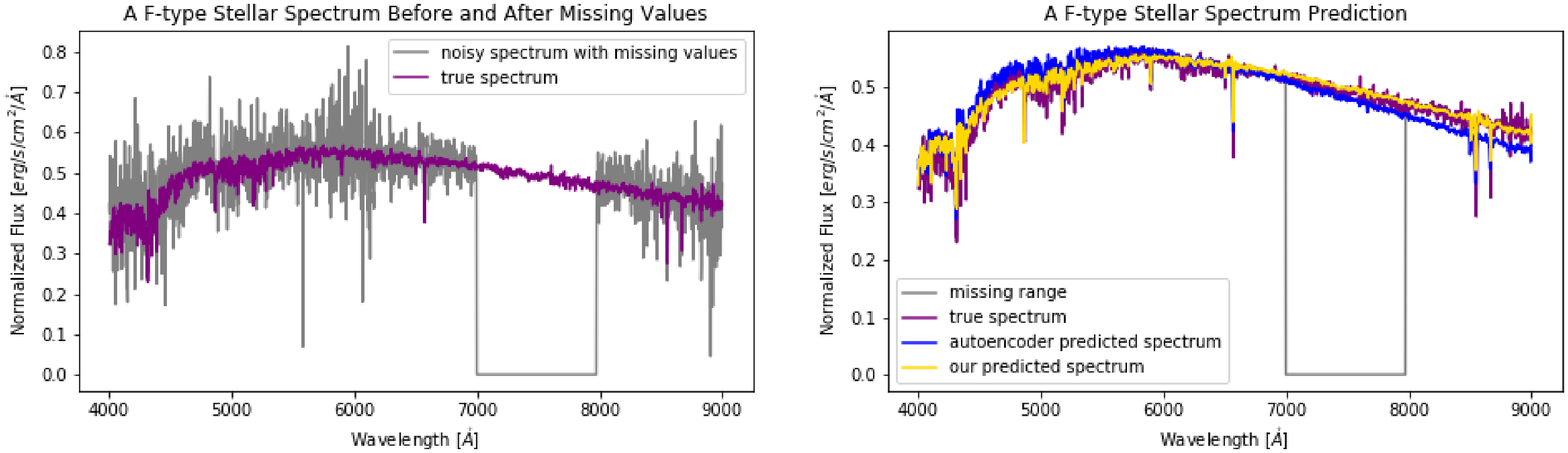}
	\includegraphics[scale=0.4]{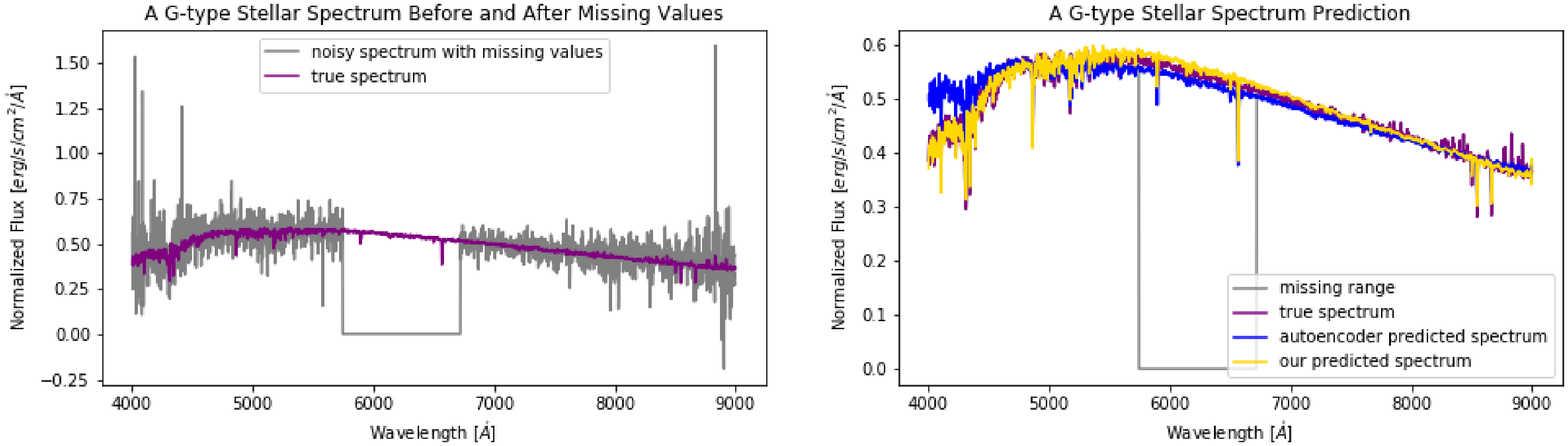}
	\caption{
		The five rows show examples of synthetic stellar spectra with missing values and the denoising results for the stellar subclasses O, B, A, F, G, respectively. The left column shows the true spectrum and its synthetic counterpart with noise and missing values. The purple curve is the clean spectrum that both denoising algorithms try to recover, and the grey curve is the  spectrum added with  noise and missing values. The  spectra denoised by both algorithms are compared  in the right column. The denoised spectrum from our model (yellow curve) is much closer to the purple clean spectrum than the standard auto-encoder result (blue curve).
	 \label{fig:testing3}}
\end{figure}

\begin{figure}
	\centering
	\includegraphics[scale=0.4]{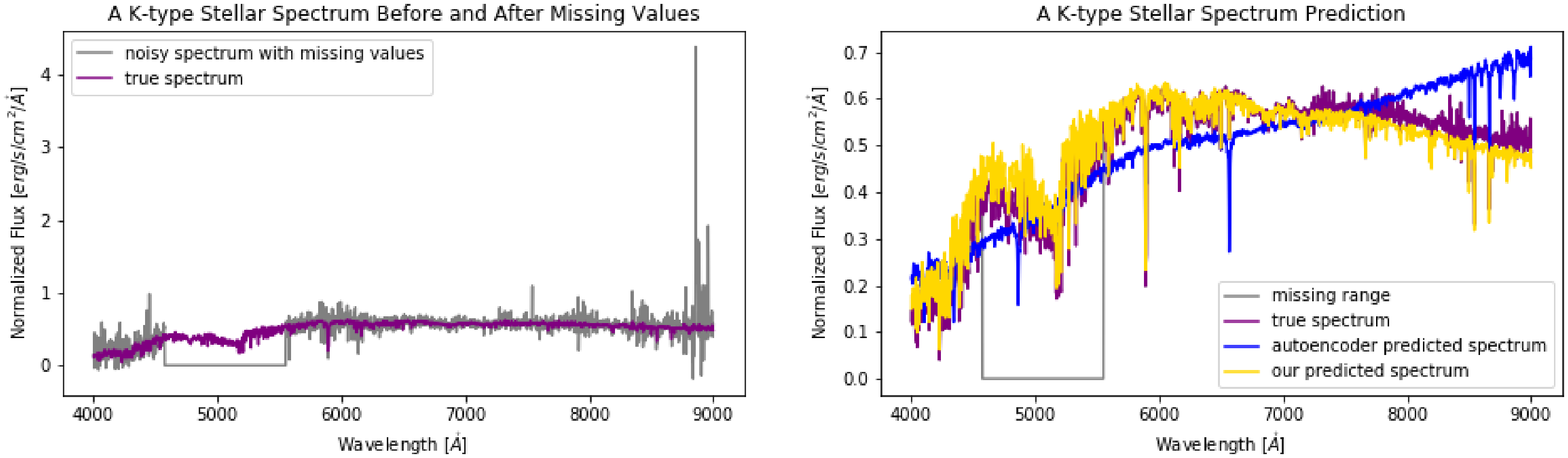}
	\includegraphics[scale=0.4]{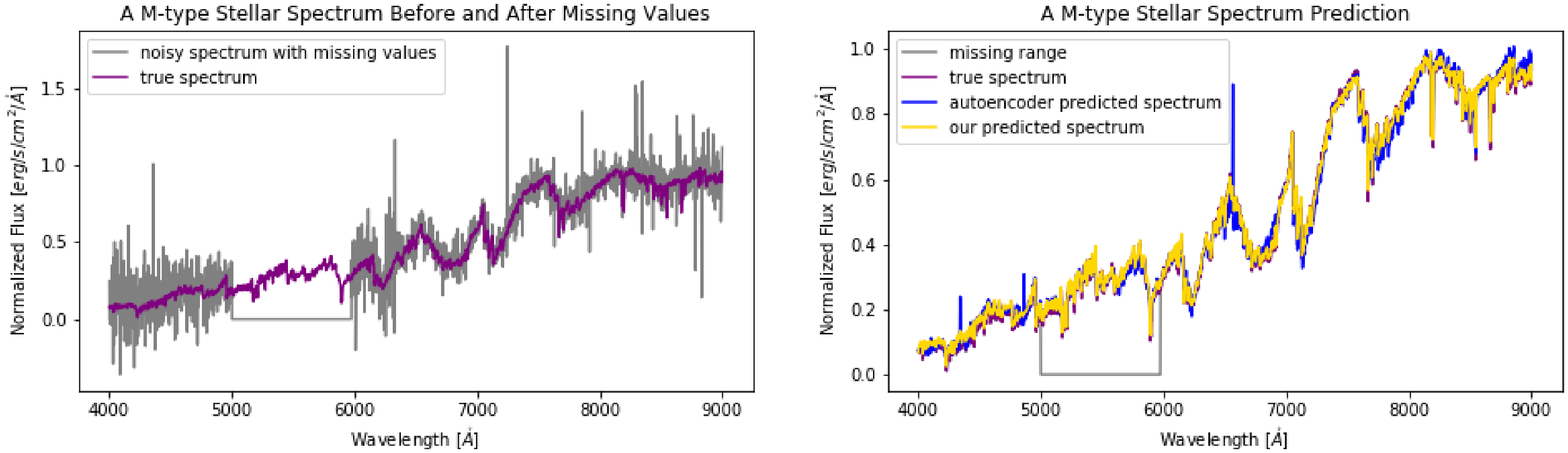}
	\includegraphics[scale=0.4]{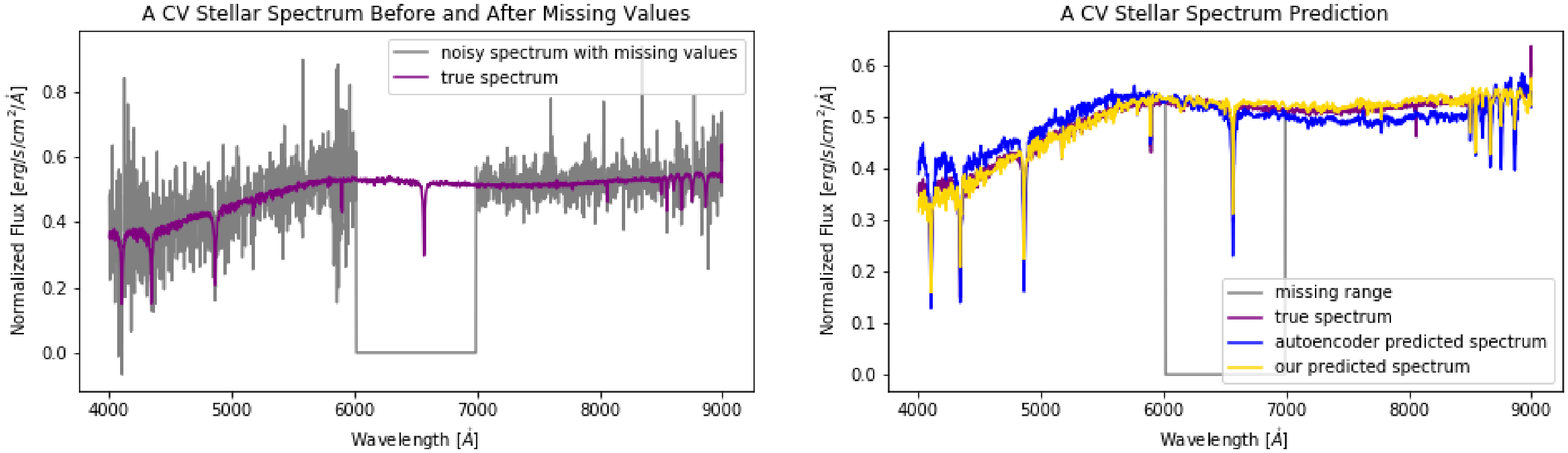}
	\includegraphics[scale=0.4]{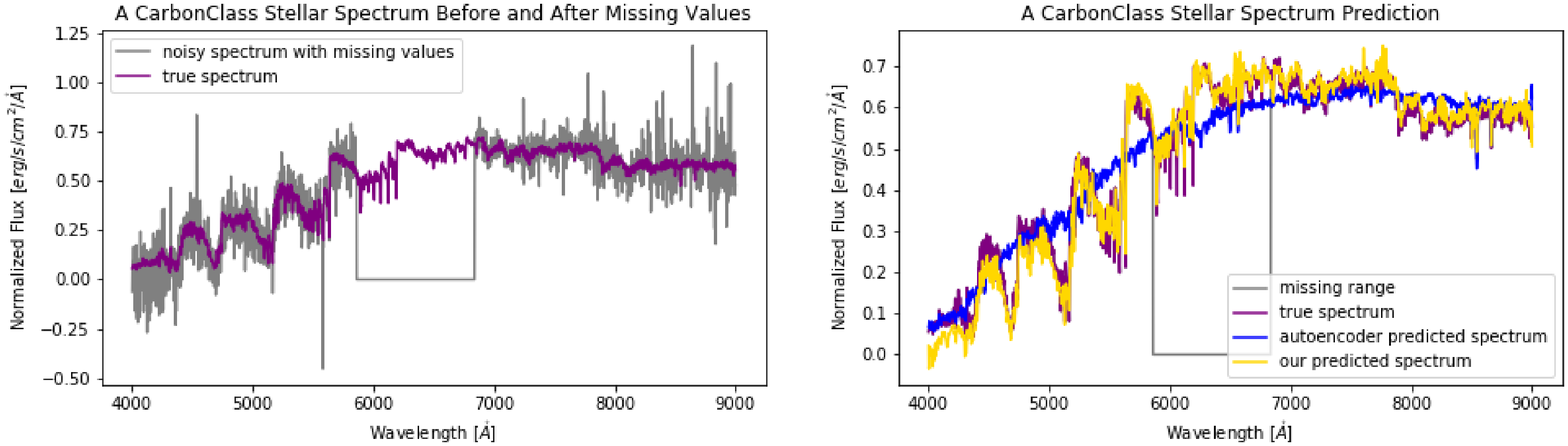}
	\includegraphics[scale=0.4]{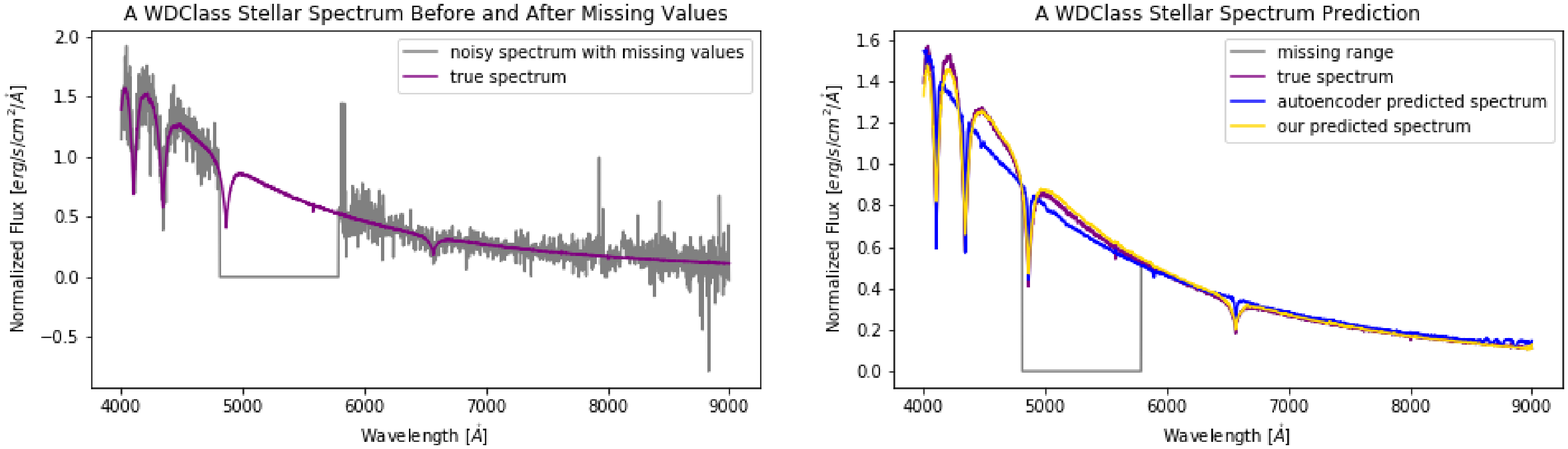}
	\caption{The five rows show examples of synthetic stellar spectra with missing values and the denoising results for the stellar subclasses K, M, CV, Carbon, WD, respectively. The left column shows the true spectrum and its synthetic counterpart with noise and missing values. The purple curve is the clean spectrum that both denoising algorithms try to recover, and the grey curve is the  spectrum added with  noise and missing values. The  spectra denoised by both algorithms are compared  in the right column. The denoised spectrum from our model (yellow curve) is much closer to the purple clean spectrum than the standard auto-encoder result (blue curve).
\label{fig:testing4}}
\end{figure}

\begin{figure}
	\centering
    \begin{minipage}{0.49\linewidth}
        \includegraphics[scale=0.42]{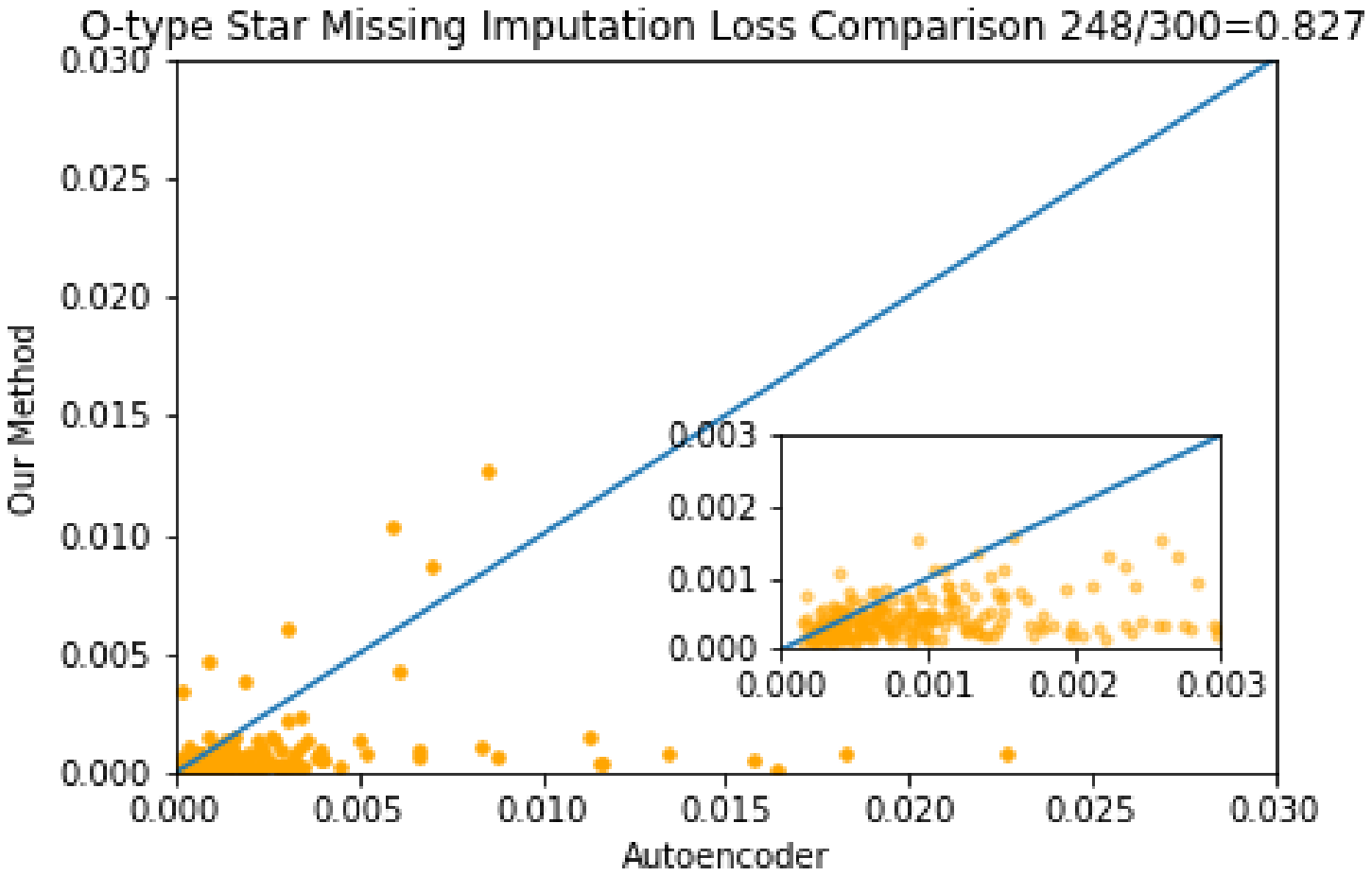}
    \end{minipage}
    \begin{minipage}{0.49\linewidth}
        \includegraphics[scale=0.42]{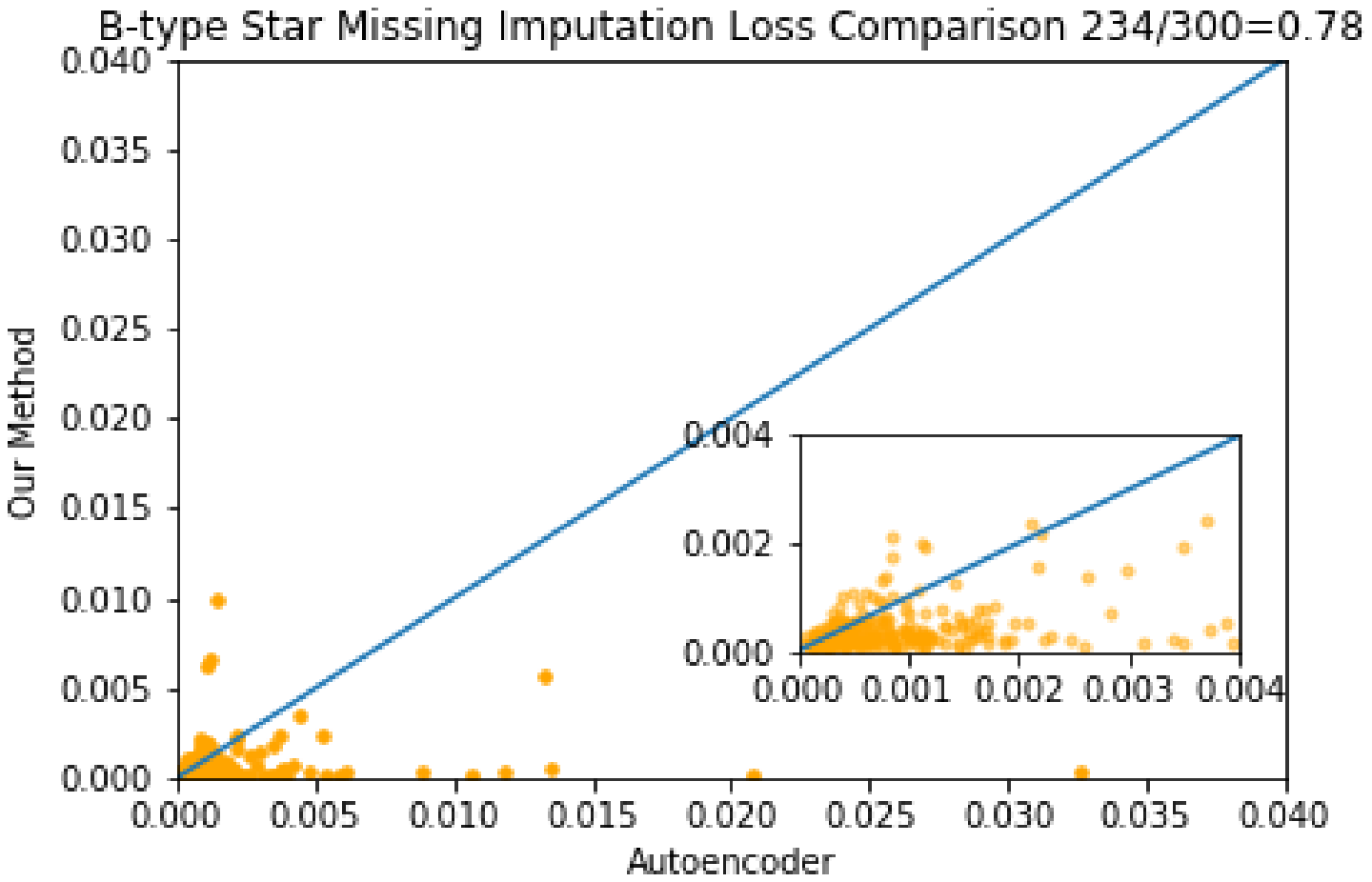}
    \end{minipage}
    \begin{minipage}{0.49\linewidth}
        \includegraphics[scale=0.42]{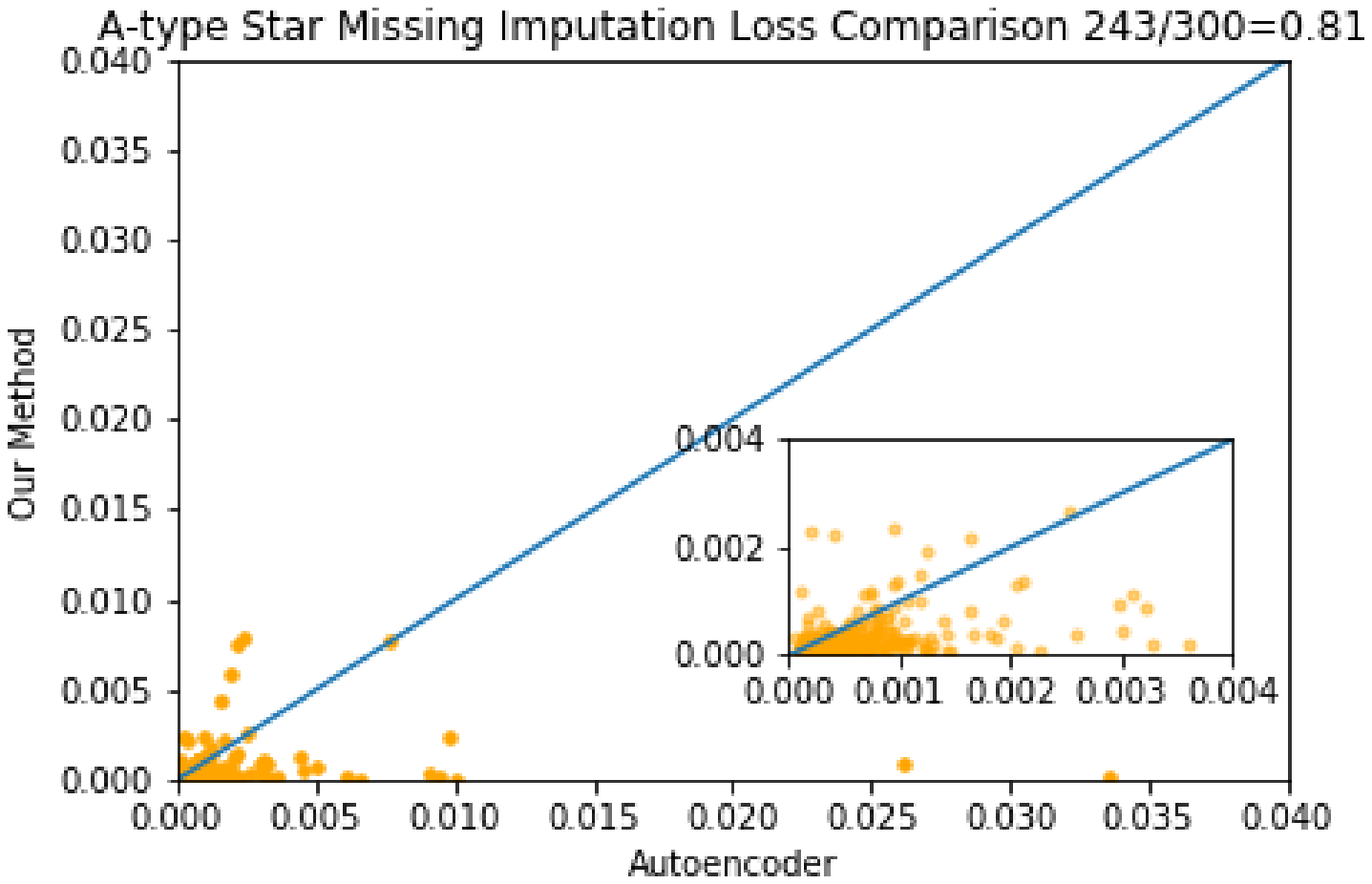}
    \end{minipage}
    \begin{minipage}{0.49\linewidth}
        \includegraphics[scale=0.42]{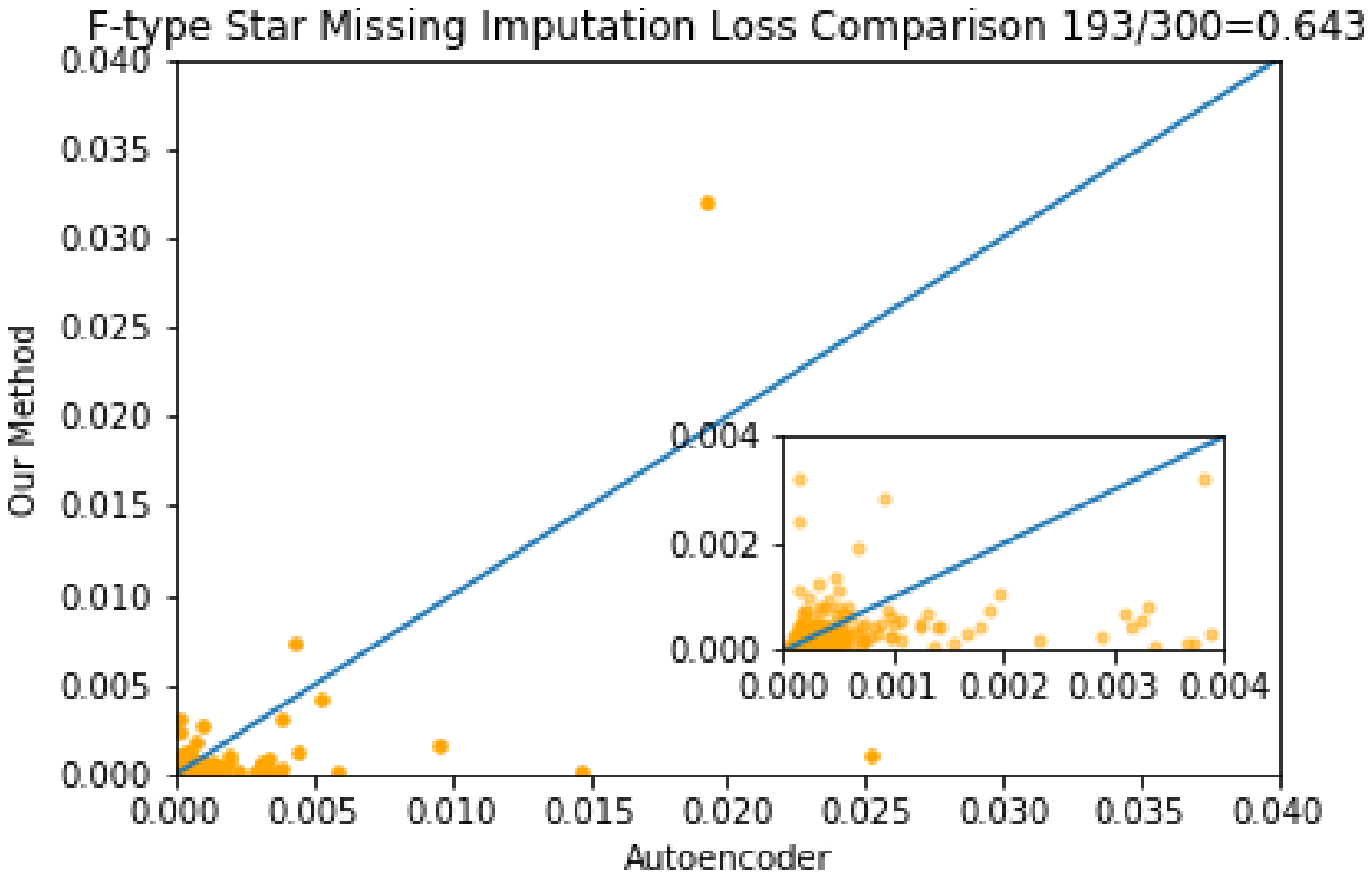}
    \end{minipage}
     \begin{minipage}{0.49\linewidth}
        \includegraphics[scale=0.42]{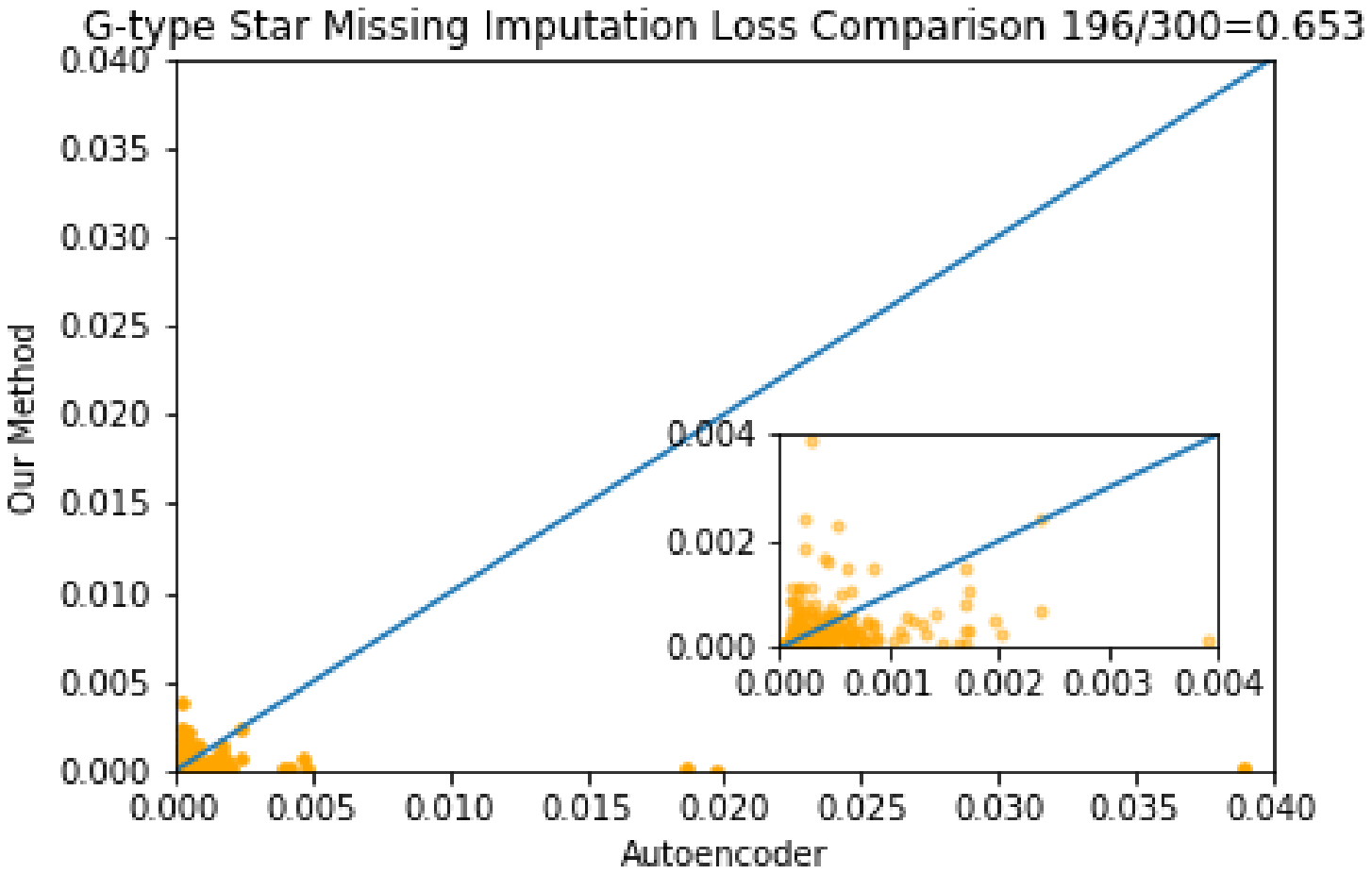}
    \end{minipage}
    \begin{minipage}{0.49\linewidth}
        \includegraphics[scale=0.42]{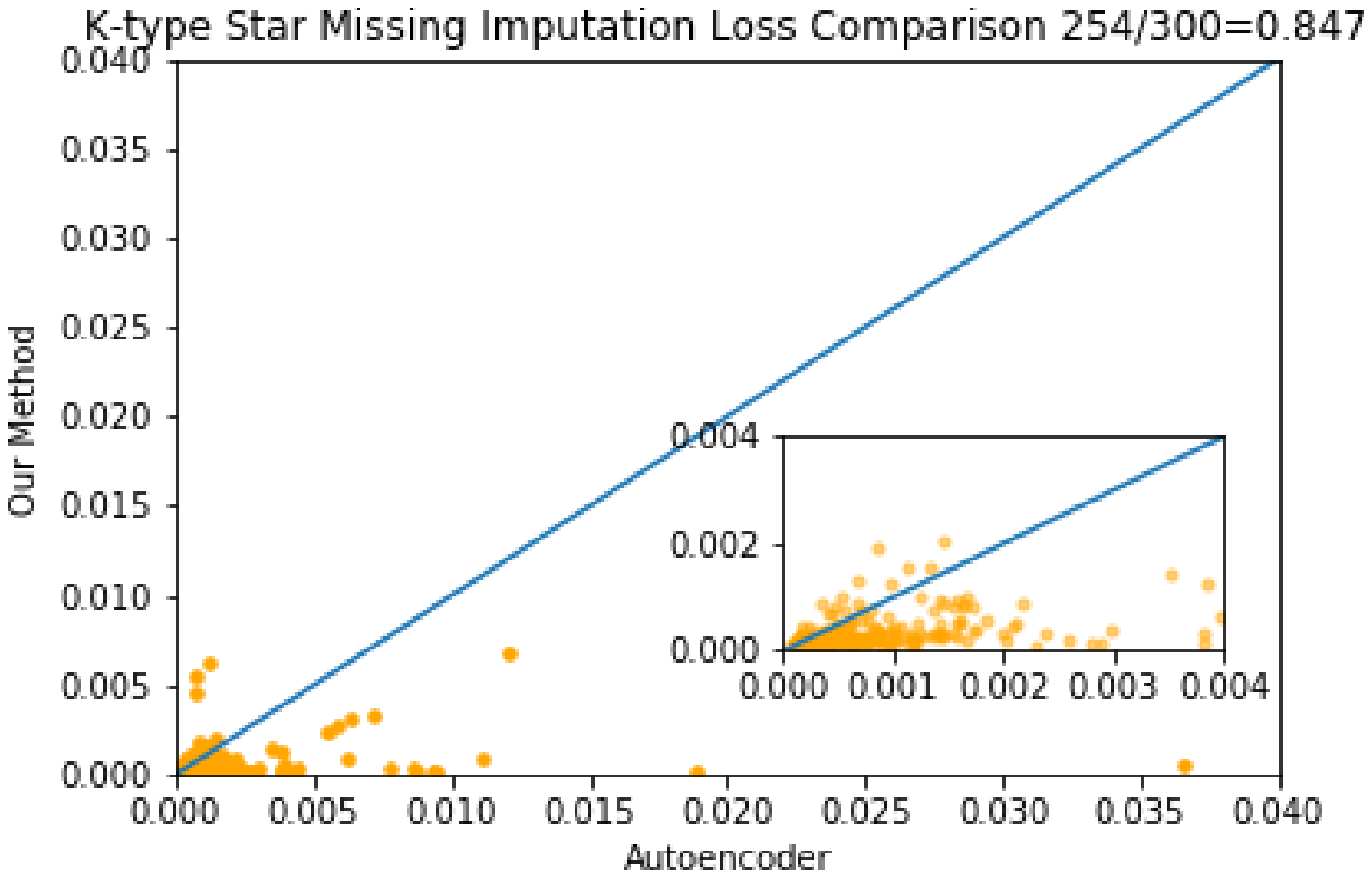}
    \end{minipage}
     \begin{minipage}{0.49\linewidth}
        \includegraphics[scale=0.42]{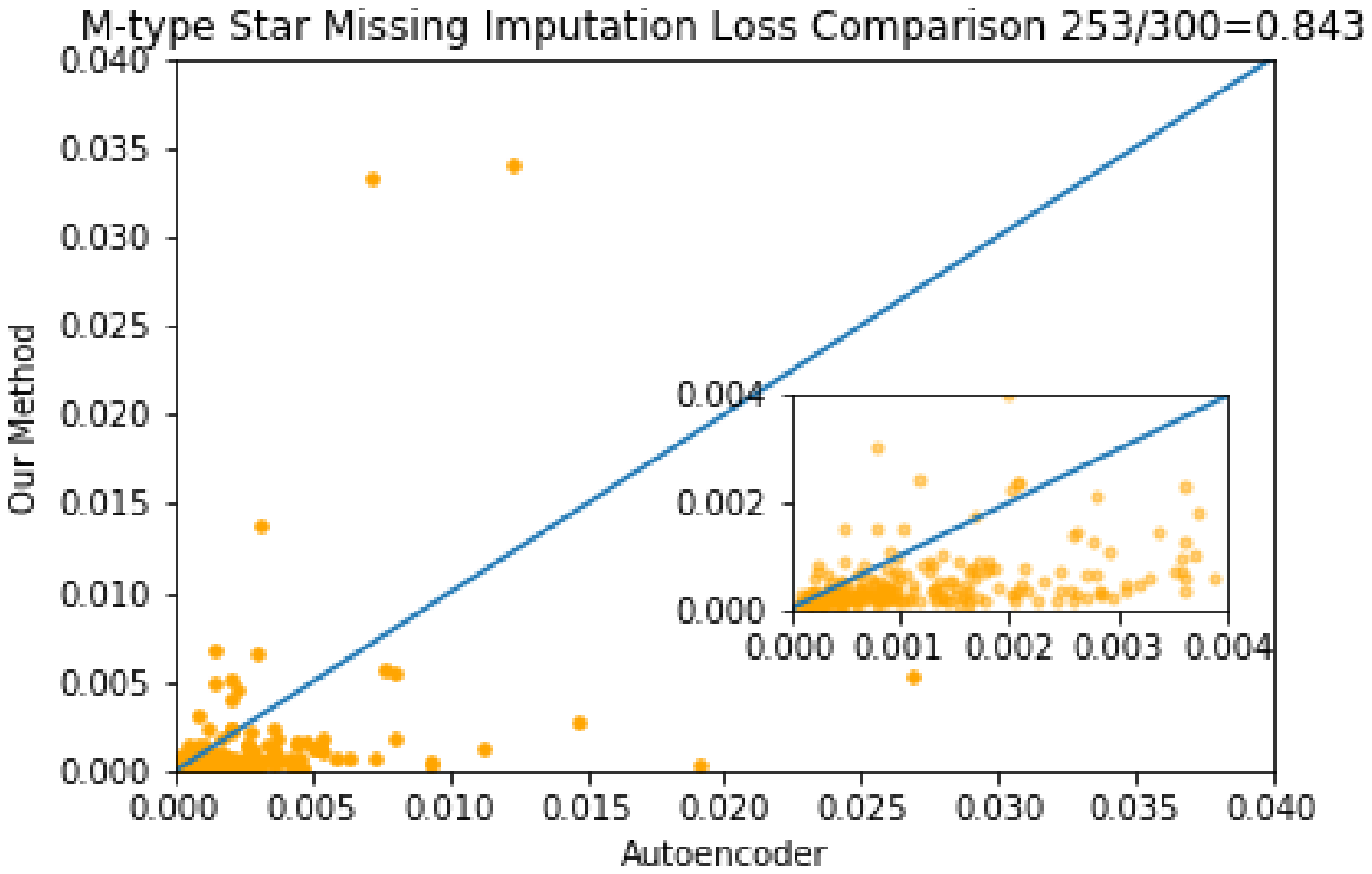}
    \end{minipage}
    \begin{minipage}{0.49\linewidth}
        \includegraphics[scale=0.42]{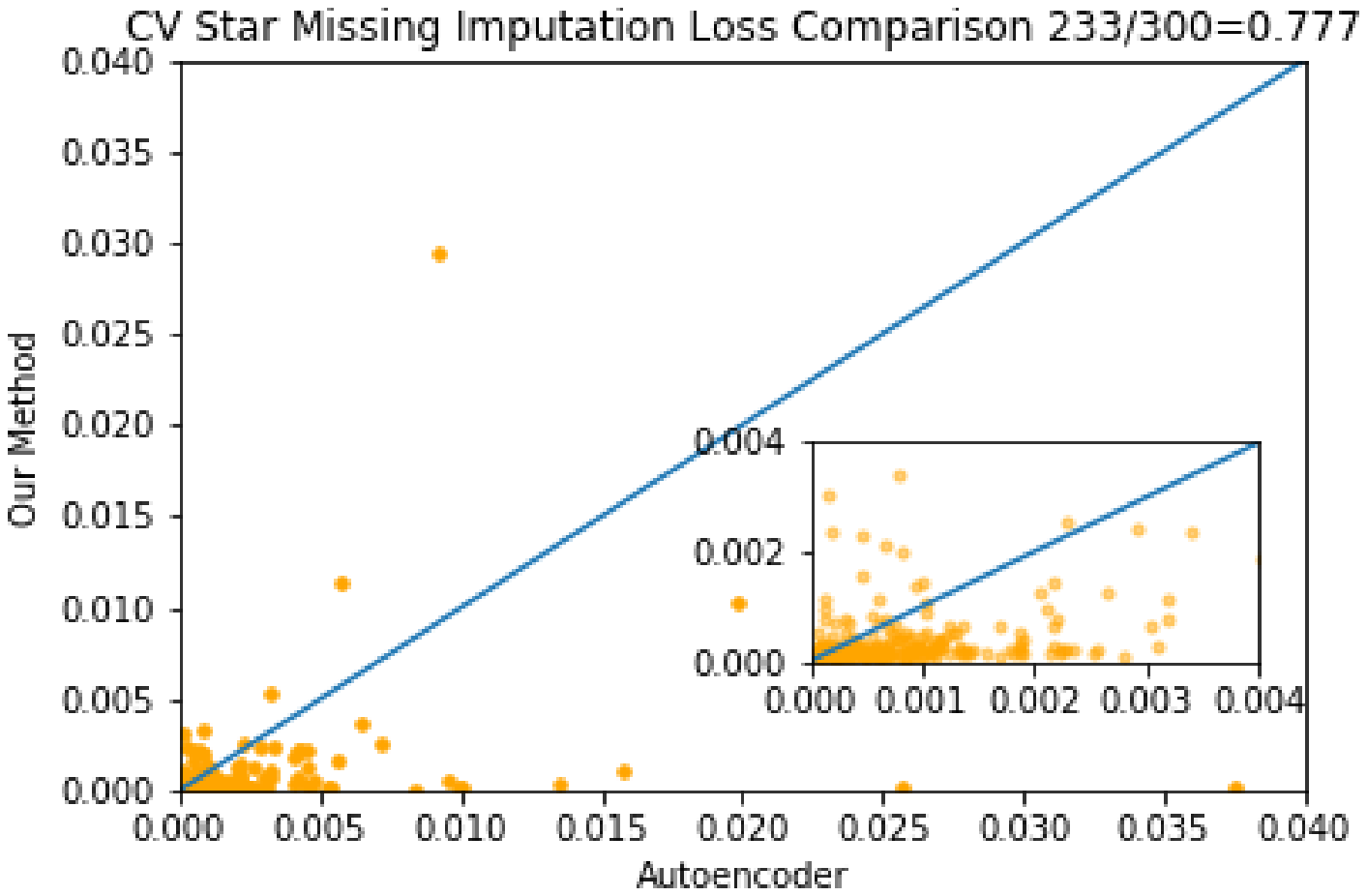}
    \end{minipage}
     \begin{minipage}{0.49\linewidth}
        \includegraphics[scale=0.42]{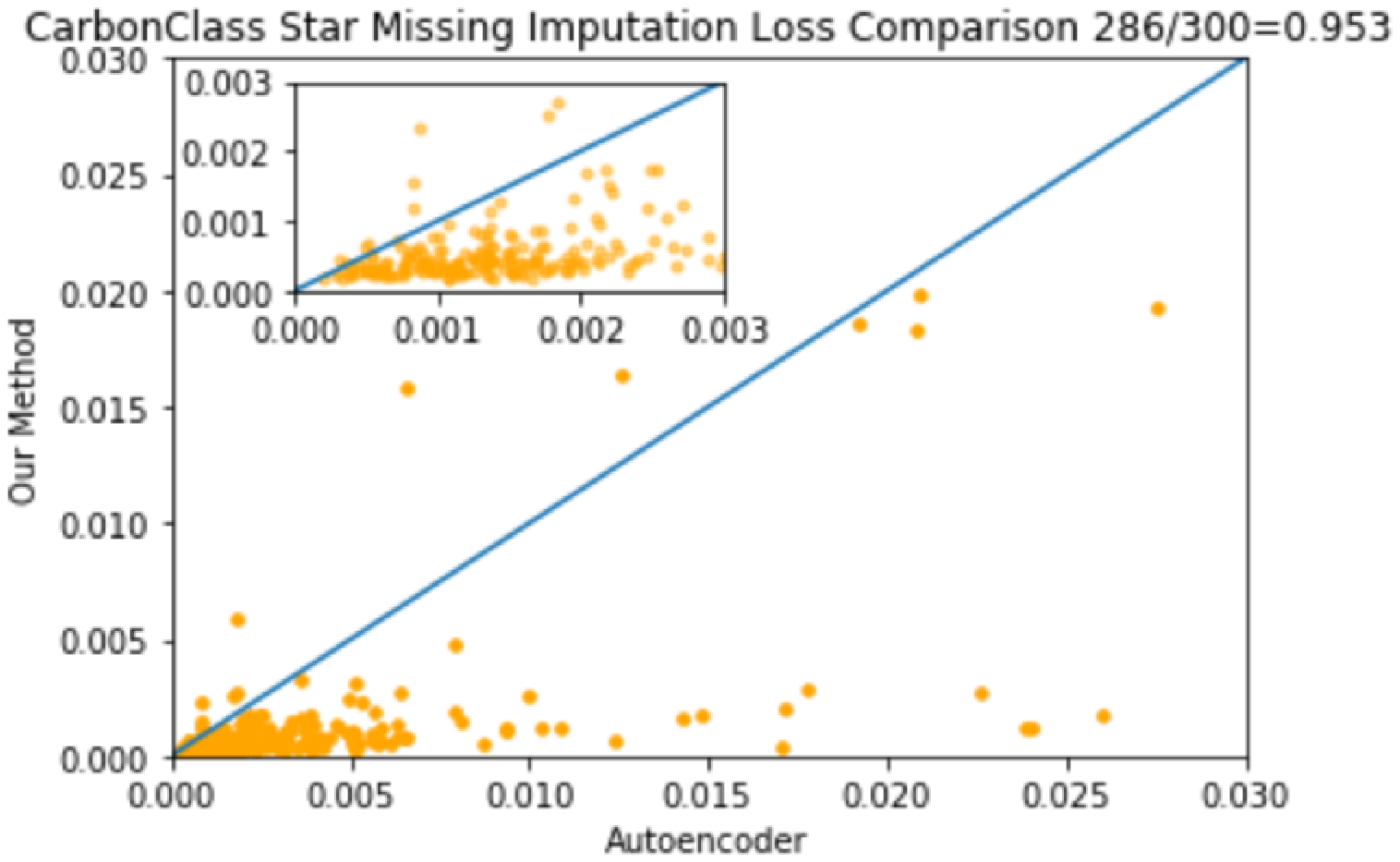}
    \end{minipage}
    \begin{minipage}{0.49\linewidth}
        \includegraphics[scale=0.42]{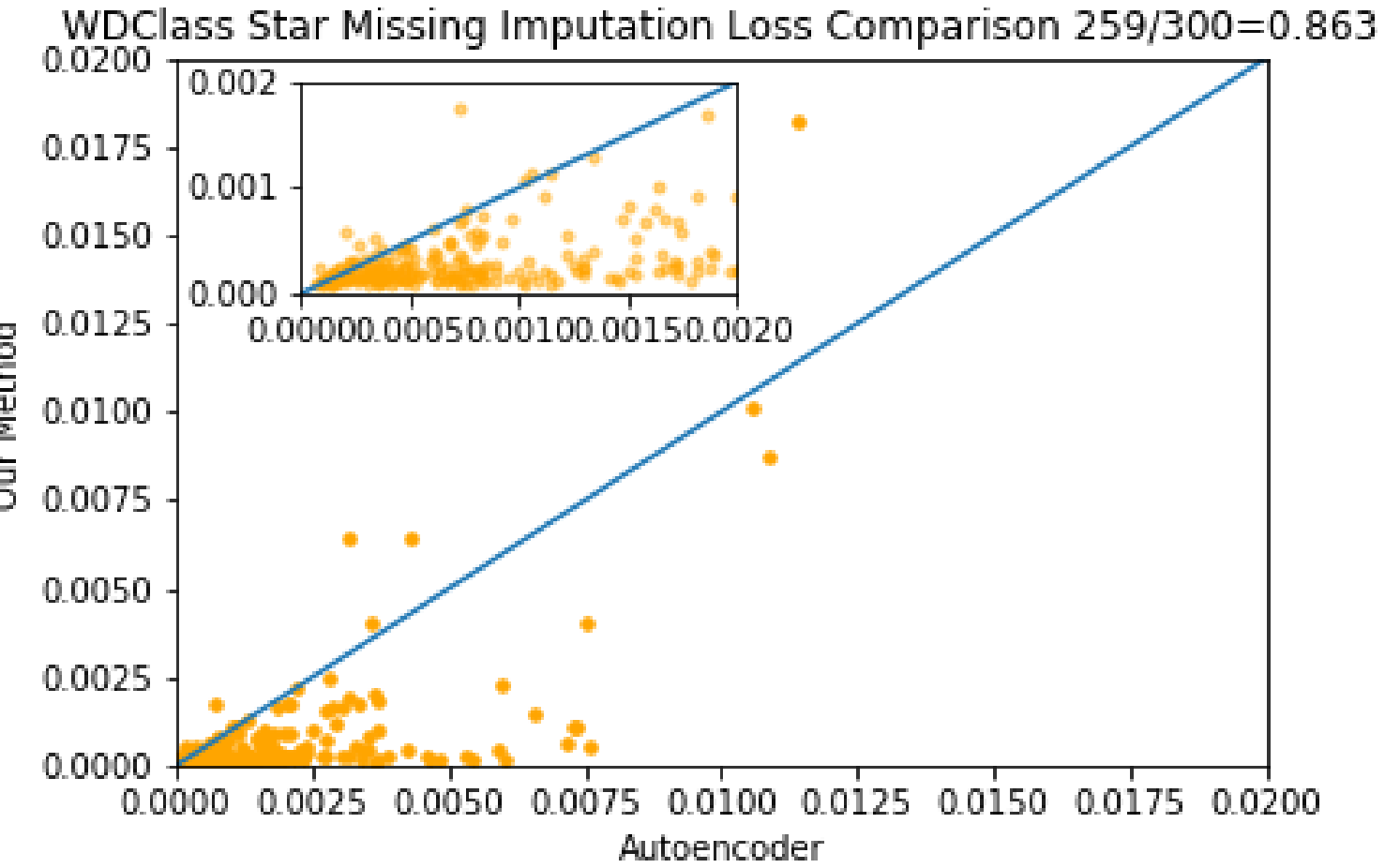}
    \end{minipage}
	\caption{
The loss comparison for the two methods applied to the noisy spectra with missing values. Each panel corresponds to one stellar subclasse. The horizontal axis is the reconstruction loss for the convolutional denoising auto-encoder, and the vertical axis is the reconstruction loss for our method. Each yellow point in a subfigure corresponds to one synthetic noisy spectrum with missing values. The blue line indicates where the two methods have equal performance. Most points in each subfigure fall below the blue line, indicating that our method has smaller reconstruction loss. The title of each subfigure also reports the proportion of spectra for which our method has smaller reconstruction loss. Our method demonstrates improved performance.	}
	\label{fig:lossMissing}
\end{figure}

Figure~\ref{fig:testing3} and Figure~\ref{fig:testing4} plot a few examples for the ten stellar subclasses. The left column shows the spectrum before and after noise addition and flux value removal. The purple spectrum is the original spectrum with high SNR. The grey spectrum has noise added, but at the same time, a random interval of flux values is removed. The missing flux is plotted as an interval of zeros. The purple spectrum is the ground truth spectrum that both algorithms try to recover. The denoised and imputed spectrum is shown in the right column. Our predicted spectrum is shown in yellow, and the result of convolutional auto-encoder is shown in blue. Our resulting spectra are much closer to the true spectra in these cases. The overall result for this test data is summarized in Figure~\ref{fig:lossMissing}. The interpretation of the figure is similar to that of Figure~\ref{fig:lossDenoise}. Although our model has a moderate lead over the convolutional auto-encoder in F-type and G-type classes, the performance difference margins are wider in the other stellar subclasses.

\section{Summary and Conclusions}


In this paper, we propose a new efficient deep Bayesian model for stellar spectral denoising, defective spectral recovery and sky emission lines or cosmic rays removal. Compared with the existing methods, our model makes a greater usage of available data, exhibits a high robustness and a superior performance in spectral denoising.  In summary,  our approach has the following advantages:
\begin{enumerate}
	\item The observation model $p(\vy | \vvs )$ takes into account the noise correlation structure. It is able to properly handle the strong sky emissions, cosmic rays, and the background noise of the observational instruments.
	\item When some part of the observation is missing due to unpredictable errors (e.g. pipeline handling error, defective spectra), our model only computes the likelihood of the observed pixels, without resorting to ad-hoc missing value imputation.
	\item Our prior model $p(\vvs)$ encodes how a true signal should look like, making our model less susceptible to defective or distorted observations (due to combining the blue and red channels).
	\item Our proposed model can also directly exploit multiple-exposure data, making the posterior inference more reliable than using only one single average data.
\end{enumerate}
The proposed method can be considered as a novel model for large-scale astronomical spectral surveys and will benefit subsequent astronomical research. In future work, we will continue refining the proposed model and investigating its proper applications in other astronomical spectral analysis tasks. For example, our model will be applied during stellar spectral data preprocessing when performing stellar classification or estimating stellar physical parameters ($T_{\rm eff}$, log$g$, [Fe/H]).

\section{Acknowledgements}
The authors would like to thank an anonymous reviewer for the helpful comments to improve  the work. This paper is funded by the National Natural Science Foundation of China under grants No.11873066 and No.U1731109. We acknowledgment SDSS databases. Funding for the Sloan Digital Sky Survey IV has been provided by the Alfred P. Sloan Foundation, the U.S. Department of Energy Office of Science, and the Participating Institutions. SDSS-IV acknowledges support and resources from the Center for High-Performance Computing at the University of Utah. The SDSS web site is www.sdss.org. SDSS-IV is managed by the Astrophysical Research Consortium for the Participating Institutions of the SDSS Collaboration including the Brazilian Participation Group, the Carnegie Institution for Science, Carnegie Mellon University, the Chilean Participation Group, the French Participation Group, Harvard-Smithsonian Center for Astrophysics, Instituto de Astrof\'isica de Canarias, The Johns Hopkins University, Kavli Institute for the Physics and Mathematics of the Universe (IPMU) /University of Tokyo, Lawrence Berkeley National Laboratory, Leibniz Institut f\"ur Astrophysik Potsdam (AIP), Max-Planck-Institut f\"ur Astronomie (MPIA Heidelberg), Max-Planck-Institut f\"ur Astrophysik (MPA Garching), Max-Planck-Institut f\"ur Extraterrestrische Physik (MPE), National Astronomical Observatories of China, New Mexico State University, New York University, University of Notre Dame, Observat\'ario Nacional / MCTI, The Ohio State University, Pennsylvania State University, Shanghai Astronomical Observatory, United Kingdom Participation Group, Universidad Nacional Aut\'onoma de M\'exico, University of Arizona, University of Colorado Boulder, University of Oxford, University of Portsmouth, University of Utah, University of Virginia, University of Washington, University of Wisconsin, Vanderbilt University, and Yale University.

\bibliographystyle{chicago}
\bibliography{ref}

\newpage

\end{document}

%% file: notation.tex
\usepackage{amsmath,amsfonts,amssymb}
\usepackage{graphicx,psfrag,epsf}
\usepackage{enumerate}

\newcommand{\bbR}{\mathbb{R}}

\newcommand{\sE}{\mathcal{E}}

\newcommand{\sG}{\mathcal{G}}

\newcommand{\vh}{\mathbf{h}}

\newcommand{\vm}{\mathbf{m}}

\newcommand{\vvs}{\mathbf{s}}

\newcommand{\vx}{\mathbf{x}}
\newcommand{\vy}{\mathbf{y}}
\newcommand{\vz}{\mathbf{z}}

\newcommand{\valpha}{\boldsymbol{\alpha}}

\newcommand{\vmu}{\boldsymbol{\mu}}
\newcommand{\vdelta}{\boldsymbol{\delta}}

\newcommand{\vepsilon}{\boldsymbol{\epsilon}}

\newcommand{\Expect}{\mathbb{E}}

\DeclareMathOperator*{\argmax}{arg\,max}